\documentclass[journal,onecolumn]{IEEEtran}

\usepackage[utf8]{inputenc}

\def\BibTeX{{\rm B\kern-.05em{\sc i\kern-.025em b}\kern-.08em
    T\kern-.1667em\lower.7ex\hbox{E}\kern-.125emX}}

\usepackage{url}

\usepackage{amsmath}

\usepackage{mathtools,halloweenmath}

\makeatletter
\newcommand{\add@empty@sup}{\@ifnextchar^{}{^{}}}

\def\IDP{
\mathsf I_{\mathsf{dp}}}
\def\Ber{
\operatorname{Ber}}
\def\dB{
\delta_\mathsf{y}}
\def\dA{
\delta_\mathsf{x}}
\def\eB{
\epsilon_\mathsf{y}}
\def\eA{
\epsilon_\mathsf{x}}
\def\pA{
p_\mathsf{x}}
\def\pB{
p_\mathsf{y}}

\usepackage{soul}
\usepackage{enumerate}
\usepackage{amsmath,amssymb,bm,amsfonts}
\usepackage{graphics,graphicx,epsfig,subfigure}
\usepackage{theorem}
\usepackage{threeparttable}
\usepackage{stfloats}
\usepackage{cite}
\usepackage{multirow}
\usepackage{amsmath,bm,bbm}
\usepackage{color}
\usepackage{multirow}
\usepackage{subeqnarray}
\usepackage{cases}
\usepackage{colortbl}
\usepackage{wasysym}
\usepackage[T1]{fontenc}    % use 8-bit T1 fonts
\usepackage{bbm}
\usepackage{graphicx}
\usepackage{hyperref}   
\usepackage{subfigure}
\usepackage{url}            % simple URL typesetting
\usepackage{booktabs}       % professional-quality tables
\usepackage{amsfonts}       % blackboard math symbols
\usepackage{nicefrac}       % compact symbols for 1/2, etc.
\usepackage{microtype}      % microtypography
\usepackage{amsmath}
\usepackage{amssymb}
\usepackage{algorithmic} 
\usepackage[ruled,linesnumbered,vlined]{algorithm2e}

\usepackage{enumitem}
\newlist{Properties}{enumerate}{2}
\setlist[Properties]{label=Property \arabic*.,itemindent=*}

\usepackage{mathtools}

\usepackage{adjustbox}
\usepackage{cuted}
\usepackage{xcolor}
\usepackage{enumitem,kantlipsum}
\usepackage{diagbox}
\definecolor{myred}{RGB}{234, 107, 102}
\definecolor{mypur}{RGB}{166, 128, 184}

\definecolor{Blue1}{RGB}{240,248,255} %背景色浅一点的
\definecolor{Blue2}{RGB}{70,130,180}%文字颜色红一点的

\usepackage[n,advantage, operators, sets, adversary, landau, probability, notions, ff, mm, primitives, events, complexity, oracles, asymptotics, keys]{cryptocode}
\DeclareMathAlphabet{\mathcal}{OMS}{cmsy}{m}{n}

\newcommand{\BlackBox}{\rule{1.5ex}{1.5ex}}  % end of proof
\newenvironment{proof}{\par\noindent{\bf Proof\ }}{\hfill\BlackBox\\[2mm]}
 
\newtheorem{theorem}{Theorem}
\newtheorem{lemma}[theorem]{Lemma} 
\newtheorem{proposition}[theorem]{Proposition}

\newtheorem{definition}[theorem]{Definition}

\providecommand{\keywords}[1]{\textbf{\textit{Index terms---}} #1}

\newcounter{MYtempeqncnt}

\begin{document}

\title{D{$\text{P}^2$}SI: Differentially-Private PSI-Analytics}

\title{DP-PSI: Private and Secure Set Intersection \\(working paper)}

\author{
    \IEEEauthorblockN{
    Jian Du\IEEEauthorrefmark{1}, 
    Tianxi Ji\IEEEauthorrefmark{2}, 
    Jamie Cui\IEEEauthorrefmark{3}
    Lei Zhang\IEEEauthorrefmark{3}
    Yufei Lu\IEEEauthorrefmark{3}
    Pu Duan\IEEEauthorrefmark{1}
    }
    
    \IEEEauthorblockA{\IEEEauthorrefmark{1}Ant Group, Sunnyvale, CA,}
     \IEEEauthorblockA{\IEEEauthorrefmark{2}Texas Tech University, Lubbock, TX,}
    \IEEEauthorblockA{\IEEEauthorrefmark{3}Ant Group, Hangzhou, China}
    
    \IEEEauthorblockA{dujianeee@gmail.com, tiji@ttu.edu, \{shanzhu.cjm, changjun.zl, yuwen.lyf, pu.d\}@antgroup.com}

}
% \author{
% {\rm Jian Du} \thanks{Jian Du is with Carnegie Mellon University (e-mail: dujianeee@gmail.com)}\\
% Ant Group
% \and
% {\rm Pu Duan}\\
% Ant Group
% \and
% {\rm Benyu Zhang}\\
% Ant Group
% }
\maketitle
\thispagestyle{plain}
\pagestyle{plain}

\begin{abstract}
One way to categorize private set intersection (PSI) for secure 2-party computation is whether the intersection is (a) revealed or (b) hidden from both parties while only computing function of the matched payload. Both aim to   provide cryptographically security while avoiding exposing each other's unmatched elements.
They, however, may be insufficient to achieve security and privacy of one practical scenario: when the intersection is required 
and
one has to consider the information leaked through the output of the function, e.g.,
"singling out each element," a GDPR concept, from either party is prohibited due to legal, ethical, and competitive reasons. Two parties, such as the advertiser and the ads supplier hold sets of users for PSI computation to reveal common users to the ads supplier in joint marketing applications, for example. In addition to the security guarantees to secure unmatched elements  required by standard PSIs, neither party is allowed to "single out" whether an element/user belongs to the other party or not, despite the fact that common users are required for joint advertising. This is an intriguing problem for which none of the PSI techniques have been able to provide a solution.
In light of this deficiency, we compose differential privacy (DP) and S2PC to give the best of both worlds, and propose {\emph {differentially-private PSI}} (DP-PSI), a new privacy model that shares PSI's strong security protection while following the recent formalization of the GDPR’s notion of singling out,
precluding  "signaling out" attack by each party except with very small probability.
The DP-PSI protocol integrates dedicated designed subsampling, shuffling, and upsampling methodology into the Diffie-Hellman (DH) PSI protocol to enable DP protection throughout the whole two-party security computation process, and makes the following benefits: 
First, while the private approximation of the PSI results are one-sided and only known by one party (e.g., the ad supplier), but they are protected from "singling out" in DP-protected states and are thus subject to data privacy rules such as the GDPR. DP-PSI also scales to billions of data points for intersection and large-scale PSI-statistics computation because the capability to be built on top of efficient elliptic curve primitives.
% $\text P^2$SI is also attractive due to its inherent privacy amplification for further statistics computation.
As a by-product, since intersection result is in the DP form, payload is not need to be sent to match the identifier for further aggregation computation, 
 DP-PSI has a much lower communication and run time overhead than the current state-of-the-art circuit-based PSI protocol.

\end{abstract}


%%%%%%%%%%%%%%%%%%%%%%%%%%%%%%%%%%%%%%%%%%%%%%%%%%%%%%%%%%%
% Main sections

% \input{sections/1_introduction}
% \newpage
% \input{sections/2_problem_formulation}
% \newpage
\section{Diffie-Hellman based PSI Preliminaries }
\label{sec:3}

% In this section, we review the security definition and classical DH based PSI.

% \subsection[Prior work: DH-PSI]{Prior work: Diffie-Hellman (DH) based PSI~\cite{DBLP:conf/sigmod/AgrawalES03}}
% \label{sec:DH}

Given an order-$p$ (prime) group $\GG$ with a group generator $g$, the DH key exchange protocol allows two parties to agree on the same key.
The protocol works as follows:
Say the sender holds a uniformly random secret key $a\sample\ZZ_p$, and the receiver holds a uniformly random secret key $b\sample\ZZ_p$, where the dollar symbol (\$) denotes that we sample an element from the distribution at random.
The sender first calculates $X^a\gets g^a\in\GG$ and sends $X^a$ to the receiver, and symmetrically, the receiver calculates $Y^b\gets g^b\in\GG$ and sends $Y^b$ to the sender.
Now, both parties are able to calculated the shared key $g^{ab}\in\GG$.
DH key exchange protocol is secure under the well-known \emph{Decisional Diffie-Hellman} (DDH) assumption.
More formally, DDH is defined as follows,

\begin{definition}
Given a DH group $\GG$ of order $p$, and $g^{a}, g^{b}$, the Decisional Diffie-Hellman (DDH) problem says that, for all $a,b\in\ZZ_p$, it is computationally hard to distinguish between $(g, g^a, g^b, g^{a b})$ and $(g, g^a, g^{b}, g^r)$, where $r\sample\ZZ_p$.
\end{definition}

In the following, we present the DH-based PSI protocol \cite{DBLP:conf/sigmod/AgrawalES03}, a classical cryptographic technique that allows two parties to compute the intersection of their data without revealing the part that is not common to the other party.
However, it does not provide privacy preserving for the intersection.
Assume there are two parties, a sender holding an input $X=[x_1, x_2,\ldots, x_n]$ and a receiver holding an input $Y=[y_1, y_2,\ldots, y_n]$ with additional payloads $\mathcal{V}=[v_{f,1}, \ldots, v_{f,n}]$.
Also, assume there is a hash function $\hash : \bin^*\to\GG$.
During the protocol setup, both parties agree on a universal Diffie-Hellman group $(\GG, m, g)$, where $m$ is the group order and $g$ is a group generator.
The sender then locally samples a secret $a\sample\ZZ_m$ and symmetrically the receiver locally samples his secret as $b\sample \ZZ_m$, where
$\$$ denotes that we sample an element from the a predefined domain at random.
After the setup, 
\begin{enumerate}
\item
  The sender first calculates and sends $[\hash(x_1)^a, \ldots, \hash(x_n)^a]$ to the receiver, while the receiver calculates and sends $[\hash(y_1)^b, \ldots, \hash(y_n)^b]$ to the sender.

\item   
    Then the sender computes $([\hash(y_1)^b, \ldots, \hash(y_n)^b])^a = [\hash(y_1)^{ab}, \ldots, \hash(y_n)^{ab}]$ and sends it to the receiver.
Meanwhile, the receiver can locally computes $([\hash(x_1)^a, \ldots, \hash(x_n)^a])^b = [\hash(x_1)^{ab}, \ldots, \hash(x_n)^{ab}]$ and calculate the intersection by finding all matched elements in $ = [\hash(x_1)^{ab}, \ldots, \hash(x_n)^{ab}]$ and $ = [\hash(y_1)^{ab}, \ldots, \hash(y_n)^{ab}]$.

\item  
Afterwards, the receiver locally calculates the aggregation of payloads if it is required $h(\{x_{f,i}\})$, with $i$ denoting all the indexes in the intersection set.
\end{enumerate}

Security for DH-PSI is formalized through the simulation paradigm, which requires the existence of simulators capable of generating views of each party in such a way that the simulated view is indistinguishable from the real protocol execution.
Specifically, if no computationally efficient algorithm can distinguish between the proposed and ideal protocols except with negligible probability.
A leakage profile $\mathcal L$~\cite{cash2013highly}, which specifies what the adversary is permitted to discover when interacting with the protocol.
That is to say, if we denote the ECC-PSI functionality as $f=(f_\mathsf{send}, f_\mathsf{recv})$ with $f_\mathsf{send}(X, Y)$ being the final output of the sender, which is the $\emptyset$ since sender receives nothing  in the ECC-PSI function output, 
and $f_\mathsf{recv}(X, Y)$ denotes the final output of the receiver, which is $X\cap Y$.
Meanwhile, the leakage profile includes $\mathcal L_{\mathsf{recv}} =|X|$ and $\mathcal L_{\mathsf{send}} =|Y|$. We rewrite the simulation based security definition for ECC-PSI as below:

% Definition 2.3 (Differential privacy for two-party protocols). We say that a protocol $P$ has $\epsilon$-differential privacy if the mechanism $\operatorname{VIEW}_{P}^{A}(x, y)$ is $\epsilon$-differentially private for all values of $x$ and same holds for $\operatorname{VIEW}_{P}^{B}(x, y)$ and all values of $y$.

% More concretely, let VIEWA
% P
% (x, y) be the joint probability distribution over x, the transcript of the
% protocol P, private randomness of the party A, where the probability space is private randomness of both
% parties. For each x, VIEWA
% P
% (x, y) is a mechanism over the y’s. Let VIEWB
% P
% (x, y) be similarly defined view
% of B whose input is y.

% \sz{Add introduction of the notations, add breif ideas about simulation security}

\begin{definition}[$\mathcal{L}$-SIM for DH-PSI]
    \label{def:sim-sec}
    DH-PSI securely realizes $f=\{f_\mathsf{send}(X,Y), f_\mathsf{recv}(X, Y)\}$ with leakage profile $\{\mathcal{L}_\mathsf{send}=|X|, \mathcal{L}_\mathsf{recv}=|Y|\}$, if for all $\lambda\in\NN^*$, $X, Y\in\bin^*$, 
    there exist two PPT simulators $\simulator_\mathsf{send}$ and $\simulator_\mathsf{recv}$ such that the following holds,
\begin{equation}
\begin{split}
    &\{\simulator_\mathsf{send}(1^\lambda, X, f_\mathsf{send}(X, Y), \mathcal{L}_\mathsf{recv})\}_{X, Y,\lambda}  
    \cindist  \{\mathsf{view}_\mathsf{send}(\lambda, X, Y)\}_{X, Y,\lambda}
    \\
    &\{\simulator_\mathsf{recv}(1^\lambda, Y, f_\mathsf{recv}(X, Y), \mathcal{L}_\mathsf{send})\}_{X, Y,\lambda}  \cindist\{(\mathsf{view}_\mathsf{recv}(\lambda, X, Y)\}_{X, Y,\lambda}.
\end{split}
\end{equation}

\noindent

\end{definition}

Note that the above security definition captures the requirement that ``nothing is learned from the protocol'' except
$f_\mathsf{send}(X, Y)$ and $f_\mathsf{recv}(X, Y)$
by showing the indistinguishability of the simulated view and the real view  with $\cindist$ to denote \textit{computational indistinguishability}.
However, since $f_\mathsf{recv}(X, Y)=X\cap Y\subseteq Y$, it does not provide  privacy preservation for $Y$.
We propose DP-PSI in this study to safeguard data privacy throughout the protocol, including the final intersection findings.

% the security of differentially private PSI is also usually defined in a simulated-based fashion \cite{groce2019cheaper}.
% % 
% Roughly speaking, their definition is parameterized with a differentially private leakage profile $\mathcal{L}=(\mathcal{L}_\mathsf{alice}, \mathcal{L}_\mathsf{bob})$.
% % 

%%DP
\section{DP-PSI Protocol and its  Privacy Definition}
Following the security definition in the last section, we study 
Differential Privacy (DP)~\cite{dwork2006calibrating} is widely regarded as the gold standard for limiting and quantifying sensitive data privacy leakage during learning tasks. Even with access to arbitrary side information, DP prevents an adversary from confidently drawing any conclusions about whether some user's data was used. The formal definition of DP is given below.
\begin{definition}[$(\epsilon, \delta)$-DP]
    A randomized algorithm $\mathcal{M}$ satisfies $(\epsilon, \delta)$-differential privacy ($(\epsilon, \delta)$-DP), if for all neighboring inputs pair $(X, X^{\prime})$, and for any $T$ belongs to the output space of $\mathcal M$, the following holds,
    \begin{align*}
        \Pr\left[\mathcal{M}(X) \in T\right] \leq e^{\epsilon} \cdot \Pr\left[\mathcal{M}(X^{\prime}) \in T\right]+\delta,
    \end{align*}
where $\epsilon>0$ and $0<\delta<1$, and $T$ denotes the output space of the $\mathcal{M}$.
\end{definition}

Lower $\epsilon$ values indicate better privacy protection. The value $\delta$ can be interpreted as the likelihood of not achieving DP.  When $\dA=0$, the above definition reduces to the $\epsilon$-DP. Intuitively, it implies that we cannot tell whether $\mathcal M$ was run on $X$ or $X^{\prime}$ based on the output results.  An important consideration when using DP is the precise condition under which $X$ and $X^{\prime}$ are considered to be neighbors. There are two natural choices~\cite{kifer2011no} that lead to \textit{unbounded DP} and \textit{bounded DP}. 
If $X$ can be obtained from $X^\prime$ by adding or removing one element, then $X$ and $X^\prime$ are neighbors in unbounded DP; On the other hand, if $X$ can be produced from $X^\prime$ by substituting one element in $X$ with another element, then $X$ and $X^\prime$ are neighbors in bounded DP.
We focus on bounded DP in this paper, so that publishing the exact number of elements in the input dataset that satisfy $0$-DP. 
We define the ideal functionality for DP-PSI as shown in Fig.~\ref{fig:ideal-func-dpsi} with DP guarantee results for the intersection. In the following, we propose the DP-PSI protocol that guarantee the computation indistinguishable security as the ideal functionality.
\begin{figure}[h]
    \centering
    \input{dpsi}
    \caption{Ideal functionality $\mathcal{F}_\mathsf{dp}= (f_\mathsf{send}, f_\mathsf{recv})$ computed by a trusted third party. In this case, $f_\mathsf{send}= \emptyset $ and $f_\mathsf{recv}=\mathsf I_{\mathsf {dp}}$ }
    \label{fig:ideal-func-dpsi}
\end{figure}

\subsection{DP-PSI Protocol}
Similar to the DH-PSI, we assume the existence of a random oracle $\hash : \bin^*\to\GG$.
During the protocol setup, both parties agree on a Diffie-Hellman group with parameters $(\GG, p, g)$, where the sender samples a random secret $a\sample\ZZ_p$ and the receiver samples his random secret $b\sample \ZZ_p$.
After the protocol setup,

\begin{enumerate}
% \begin{list}{\labelitemi}{\leftmargin=0em}
\item 
The sender {\it reorder and encrypts} each item $x_i\in X$  by $\mathsf{HR}(x_i)^a$  and sends $X^a \triangleq [\hash(x_1)^a, \ldots, \hash(x_n)^a]$ to the receiver.

    \begin{figure}[htb]
    % \label{fig:input-output}
    \centering
    \begin{bbrenv}{A}
    \begin{bbrbox}[name=(sender)]
    encryption
    \end{bbrbox}
    \bbrmsgto{side={$X$}}
    \bbrqryto {side={$X^{a}$}, islast}
    \end{bbrenv}
    \end{figure}
\item
\begin{itemize}
    \item 
    The receiver  {\it sub-samples}  his items  in $Y$ with $\mathsf {Ber}(Y;\pB)$
     where the selection  of each item  is subjected to an independent Bernoulli trial 
    with probability $\pB$. The selected random subset is  $\mathsf{Y_{\mathsf{sub}}} \subseteq Y$.
  After the subsampling, the receiver  {\it encrypts} $\mathsf{Y_{\mathsf{sub}}}$  by pushing it into $\mathsf Y^b_{\mathsf{sub}}\gets\hash(\mathsf Y_{\mathsf{sub}})^b$ with 
  $\mathsf Y_{\mathsf{sub}}^b  \triangleq [\ldots, \hash(y_i)^b, \ldots] $ for $y_i\in \mathsf Y_{\mathsf{sub}}$.
 Then  $\mathsf Y^b_{\mathsf{sub}}$ is sent to the sender.
 
 \item
  On receiving $X^a$, the receiver samples a uniformly random permutation $\pi\sample S_{n}$,
 which is a bijective function
  with $\pi: [n]\to [n]$, and then {\it permutes} the order of items in $X^a$ by $\pi$ leading to output $X ^{a}_{\pi} \triangleq \left[\hash(x_{\pi(1)})^{a },\ldots \hash(x_{\pi(n)})^{a }\right]$.
Afterwards, the receiver {\it re-encrypts} $X_{\pi}^{a}$ to  $X_{\pi}^{ab} \triangleq   \left[\hash(x_{\pi_a(1)})^{a b},\ldots \hash(x_{\pi_a(n)})^{a b}\right]$ and sends it back to the sender.

    
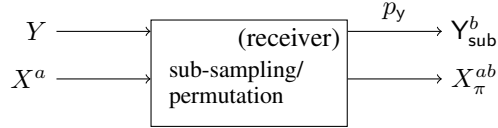
\begin{figure}[htb]
    % \label{fig:input-output}
    \centering
    \begin{bbrenv}{X}
    \begin{bbrbox}[name=(receiver)]
    \small{sub-sampling/\\permutation}
    \end{bbrbox}
    \bbrmsgto{side={$Y$}}
    \bbrmsgto{side={$X^a$}}
    \bbrqryto{top={$\pB$}, side={$\mathsf Y^{b}_{\mathsf{sub}}$}}
    \bbrqryto{ side={$X_\pi^{ab}$}, }
    \end{bbrenv}
    \end{figure}
 \end{itemize}

    \item Upon receiving $\mathsf Y^{b}_\mathsf{sub}$ and $X^{a,b}_{\pi}$, the sender  re-encrypts $\mathsf Y^{b}_\mathsf{sub}$ by $\mathsf Y^{a,b}_\mathsf{sub}\triangleq(\mathsf Y^{b}_\mathsf{sub})^a$ in order to compute the random set intersection and subtraction: 
    \begin{equation}
    \begin{split}
    % \begin{align*}
       &\text{intersection:} \quad  \mathsf I^{ab}_\mathsf{sub}
        \triangleq X^{a,b}_{\pi}\cap {\mathsf Y^{a,b}_\mathsf{sub}}.\\
       & \text{subtraction: }\quad 
       {\mathsf Y^{a,b}_\mathsf{sub}}
       \setminus
       X^{a,b}_{\pi}.
        \end{split}
    % \end{align*}
    \end{equation}
    The sender further conducts up-sampling by  adding items from  $\mathsf Y_{\mathsf{sub}}\setminus
       X^{ab}_{\pi}$ to $\mathsf I^{ab}_\mathsf{sub}$  at random with each item chosen by an independent Bernoulli trial with probability $q$, i.e.,
    \begin{equation}
        \text{upsampling:}\quad 
        {\IDP^{ab}}\triangleq { \mathsf{Ber}(\mathsf Y^{a,b}_\mathsf{sub}\setminus   \mathsf I^{ab}_\mathsf{sub}}; q)\cup   \Ber(\mathsf I_\mathsf{sub}^{ab};\pA).
    \end{equation}

    In this step, the sender preserves the indices of  items of $\IDP^{ab}$ in $\mathsf Y_\mathsf{sub}^{a,b}$,  i.e.,  ${\mathsf{Idx}(\IDP^{ab}, \mathsf Y_\mathsf{sub}^{a,b})}$ and send them to the receiver.
    
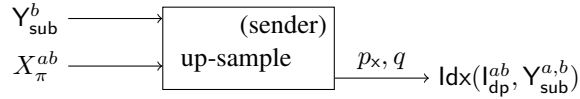
\begin{figure}[htb]
    \centering
    \begin{bbrenv}{X}
    \begin{bbrbox}[name=(sender)]
    up-sample
    \end{bbrbox}
    \bbrmsgto{side={$\mathsf Y^{b}_{\mathsf{sub}}$}}
    \bbrmsgto{side={$X_\pi^{ab}$}}
    \bbrqryto{top={$\pA, q$}, side={$\mathsf{Idx}(\IDP^{ab}, \mathsf Y_\mathsf{sub}^{a,b})$}, islast}
    \end{bbrenv}
    \end{figure}
    \item Finally, the receiver is able to identify $\mathsf {I_{DP}}$, the
    DP assured PSI result,  by taking elements from ${\mathsf Y_{\mathsf{sub}}}$ with   received indices ${\mathsf{Idx}(\mathsf I_{\mathsf{DP}}^{ab}, \mathsf Y_\mathsf{sub}^{a,b})}$.
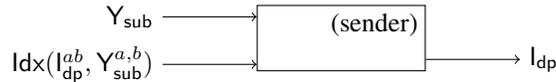
        \begin{figure}[htb]
    \centering
    \begin{bbrenv}{X}
    \begin{bbrbox}[name=(sender)]
     
    \end{bbrbox}
    \bbrmsgto{side={$\mathsf Y_{\mathsf{sub}}$}}
    \bbrmsgto{side={$\mathsf{Idx}(\IDP^{ab}, \mathsf Y_\mathsf{sub}^{a,b})$}}
    \bbrqryto{ side={$\IDP$}, islast}
    \end{bbrenv}
    \end{figure}
\end{enumerate}
% \end{list}
We summarize the above DP-PSI protocol in Fig.~\ref{fig:dp-psi}.
\noindent Remark:
\begin{enumerate}
    \item   
    Step 1) is identical to the original DH-PSI protocol.
    Since the receiver lacks the decryption key, the plaintexts must remain hidden, and the only piece of information the receiver  obtains is the cardinality $|X|$.
Yet, publication of $|X|$ satisfies $0$-DP in accordance with the bounded DP interpretation, according to the its definition~\cite[p.~150]{li2016differential}. 
\item
Here, we observe that all elements in $\mathsf I^{ab}_\mathsf{sub}$  belong to the set $\mathsf Y_{\mathsf{sub}}\subseteq Y$. And $\mathsf I^{ab}_\mathsf{sub}$  is a probablistic subset of  due to the sub-sampling process of $\mathsf Y_{\mathsf{sub}}$ at the receiver. 
\item
In this protocol, we have the DP leakage profile:
\begin{equation}
    \mathcal{M}_\mathsf{recv}(Y) =\{|\mathsf Y_{\mathsf{sub}}|,  |X\cap \mathsf Y_{\mathsf{sub}}| \},
\quad
    \mathcal{M}_\mathsf{send}(X) =\{|X|\}.
\end{equation} 
\end{enumerate}

More concretely, we define the ideal functionality at sender and receiver side as $f_\mathsf{send}$ and $f_\mathsf{recv}$, respectively. According to Fig \ref{fig:ideal-func-dpsi}, we have $f_\mathsf{send}=\emptyset$ and $f_\mathsf{recv}=\IDP$. Also, since we restrict the leakage satisfy the definition of DP, we formulate the leakage profile as $\mathcal{M}_\mathsf{recv}(Y)$ as the DP leakage from receiver, and $\mathcal{M}_\mathsf{send}(X)$ as the DP leakage from sender. Now we have the security and privacy  definition for DP-PSI as following.

% \hl{todo: shanzhu: how we modify definition 2 to definition 3}

\begin{definition}[$ \mathcal{L}_{(\epsilon,\delta)}$-SIM]
    \label{def:dp-sec}
    The DP-PSI protocol  securely realizes  the ideal functionality $\mathcal{F}_\mathsf{dpsi}=(f_\mathsf{send}, f_\mathsf{recv})$ with leakage profile $\mathcal{M}_\mathsf{send}(X) $  and $ \mathcal{M}_\mathsf{recv}(Y)$, 
    if all $\lambda\in\NN^*$, $X, Y\in\bin^*$:
    \begin{enumerate}
        \item
    There exist two PPT simulators $\simulator_\mathsf{send}$ and $\simulator_\mathsf{recv}$ with the following indistinguishability holds,
\begin{align*}
    \{\simulator_\mathsf{send}(1^\lambda, X, f_\mathsf{send}(X, Y), \mathcal{M}_\mathsf{recv}(Y))\}_{X, Y,\lambda}  \cindist
    \{\mathsf{view}_\mathsf{send}(\lambda, X, Y)\}_{X, Y,\lambda}
    \\
    \{\simulator_\mathsf{recv}(1^\lambda, Y, f_\mathsf{recv}(X, Y), \mathcal{M}_\mathsf{send}(X))\}_{X, Y,\lambda}  \cindist
    \{\mathsf{view}_\mathsf{recv}(\lambda, X, Y)\}_{X, Y,\lambda}.
\end{align*}
\item
For any neighboring input pairs, $X\sim X^{\prime}$ and $Y\sim Y^\prime$, $\mathcal M_\mathsf{send}$ and $\mathcal M_\mathsf{recv}$ satisfies $0$-DP for $X$ and $(\eB,\dB)$-DP for $Y$, respectively.
\item
Moreover, for each element in $f_{\mathsf{recv}}=\IDP$, weather it belongs to sender $X$ satisfies $(\eA,\dA)$-DP.
\end{enumerate}
\end{definition}

The metric of  $\mathcal{L}_{(\epsilon,\delta)}$-SIM provides stronger privacy guarantee than $\mathcal{L}$-SIM,
 because it requires DP guarantee to the intermediate and output leakage profile.
In the following, we design our DPSI protocol and prove that it meets the definition of $ \mathcal{L}^{(\epsilon,\delta)}_\mathsf{dp}$-SIM security.
As a result of the specified designed random mechanism, all the leakage information throughout the protocol is DP guaranteed.
Furthermore,
DP-PSI provides "plausible deniability" for each record in the PSI result because the adversary has no idea why the element in the PSI result is "true" or "false." The rigorous analysis is provided in the next section.
Before that, We prove DP-PSI Figure \ref{fig:dp-psi} is $(\epsilon, \mathcal{L})$-SIM secure as below with the proof in Appendix~\ref{app:lemma:security}.
\begin{lemma}
\label{lemma:security}
Given the DP assurance of $\mathcal M_{\mathsf{send}}(X)$ for $X$, $\mathcal M_{\mathsf{recv}}(Y)$ for $Y$, and $\mathsf I_{\mathsf{DP}}$ for $X$,  
DP-PSI is a secure instantiation of $ \mathcal{L}_{(\epsilon,\delta)}$-SIM  with $(\eA, \dA)$ and $(\eB, \dB)$ protection for $X$ and $Y$, respectively.
\end{lemma}
In the following section, we demonstrate the DP guarantees of $\mathcal M_{\mathsf{send}}(X)$ for $X$, $\mathcal M_{\mathsf{recv}}(Y)$ for $Y$, and $\mathsf I_{\mathsf{DP}}$ for $X$.

\begin{figure*}[htbp]
    \centering
    \input{main_protocol}
    \caption{The DP-PSI protocol for  computing the intersection of $X$ and $Y$  that meets the requirement for privacy by not disclosing  information about individual records in either parties.
    % Fig~\ref{fig:dp-psi-example} in the appendix is an concrete example of this DP-PSI protocol.
    }
    \label{fig:dp-psi}
\end{figure*}

% \newpage
% \input{sections/4_our_protocol}
% \clearpage
\section{DP Analysis}
As shown in Fig.~\ref{fig:dp-psi}, the entire DP-PSI protocol incorporates the sender's and the receiver's views on intermediate results.
Because the DP-PSI  requires DP assurance not only for the final intersection but also for all intermediate results, the DP analyses must be thoroughly examined independently from the sender's and the receiver's data protection perspectives.
As a result, we deconstruct the operations of the DP-PSI protocol to identify random mechanisms including $\mathcal M_X$ and $\mathcal M_Y$, which correspond to DP protection for the sender and receiver, respectively. We extract these mechanisms from DP-PSI and  conduct separate privacy analyses for each of them in the following subsections.
When there is no ambiguity, we will replace the ciphertext with its plaintext counterpart in the following privacy analysis to simplify the notation. 

% The mappings used to convert ciphertext to plaintext are summarized below, and we start from the privacy analyses for Bob.
% \begin{Properties}
%   \item 
%   $\hash(y_i)^{ab}\in X^{ab}_{\pi}\iff y_i\in X$
  
%   \item 
%   $I^{ab}_{\mathsf{sub}}=X^{ab}_{\pi}\cap Y^{ab}_\mathsf{sub}
% \iff 
% |X \cap \mathsf{Y_{sub}}|$
 
%   \item 
%   $x_i\in X\setminus B$ and $x_i\notin X\cap Y$,  implies that
% $x_i\notin I_{\mathsf{DP}}$

% \item 
% $  \textcolor{myred}{\mathsf I^{ab}_\mathsf{sub}}\triangleq X^{ab}_{\pi}\cap \textcolor{myred}{\mathsf{Y}^{ab}_\mathsf{sub}}$
% \item 
% \end{Properties}

\subsection{$\mathcal M_{\mathsf Y}$: Privacy for  $Y$ }
In Step~3) of DP-PSI, thanks to the fact that $X^{ab}_{\pi}$ is a random permutation of $X^{ab}$, the sender cannot figure out the plaintext counterpart  in $\mathsf I_{\mathsf{sub}}^{ab}=X^{ab}_{\pi}\cap Y^{ab}_\mathsf{sub}$. Therefore,
the sender cannot observe any other information about the receiver's data except for the cardinality of the intersection, $|{\mathsf I^{ab}_\mathsf{sub}}|= |X^{ab}_{\pi}\cap Y^{ab}_\mathsf{sub}|$.
In this subsection, we will investigate the DP for $Y$ given the observation of $\mathsf I^{ab}_{\mathsf{sub}}$ as shown in  Algorithm~\ref{algo:real-partical-protocol}

\begin{algorithm}
% \footnotesize
\SetKwInOut{Input}{Input}
\SetKwInOut{Output}{Output}
\Input {$X$, $Y$, $\hash(\cdot)^x$, $\pB$, $ a\sample\ZZ_p$, and $b\sample\ZZ_p$ .} %Y^{ab}_\mathsf{sub}$
\Output{$|\mathsf I^{ab}_{\mathsf{sub}}|$.}

% \ForAll{$y_i\in B$}{

% Let $\mathsf{x}_i \leftarrow \Ber\left(p\right)$

% \If{$\mathsf x_i==1$ $\land$ $y_i\in X$}{
% $\mathsf y_i\leftarrow 1$
% }
% \Else {$\mathsf y_i\leftarrow0$}
% }

% Return \hl{$ |\mathsf I^{ab}_{\mathsf{sub}}|=\sum_{y_i\in B}\mathsf y_i $.}

Order $Y$  lexicographically and then compute  $\mathsf Y_{\mathsf{sub}} =  {\mathsf{Ber}(Y;\pB)}$

Return $|\mathsf I^{ab}_{\mathsf{sub}}| = \sum_{y_i\in \mathsf Y_{\mathsf{sub}}} \mathbbm{1}[\Ber\left(\pB\right) = 1 \land y_i \in A]$

\caption{$|\mathsf I^{ab}_{\mathsf{sub}}|\triangleq \mathcal M_{\mathsf Y}(X, Y)$: The view of the sender on the cardinality of $X \cap \mathsf{Y_{sub}}$ in $\mathrm{DP^2SI}$.}   %, i.e., $\Psi \triangleq |\widetilde {\mathcal I}_{ab}|\triangleq |E_{b,a}^\mathsf{keep}\cap E^\prime_{ab}|$}
\label{algo:real-partical-protocol}
\end{algorithm}

% \begin{figure*}[htbp]
% \centering
% \includegraphics[width=1\textwidth]{images/roadma\pBob.png}
% \caption{Roadmap of constructing the privacy guarantee of Bob's data against Alice.\hl{add more description..}}
% \label{fig:roadmap-privacy}
% \end{figure*}

% It is noteworthy that under cipher text, except for the cardinality of the initial set intersection at step (4) of the protocol (i.e., ${I^{ab}_\mathsf{sub}}\triangleq X^{ab}_{\pi}\cap Y^{ab}_\mathsf{sub}$), Alice cannot observe any other information about Bob's data. This is because $X^{ab}_{\pi}$ is a random permutation of $X^{ab}$ (i.e., the cipher text of Alice's data encrypted using the private secrets of the sender and receiver).

\noindent Proof Sketch for $(\eB,\dB)$ in DP-PSI (equivalently $\mathcal M_{\mathsf Y}$ in Algorithm~\ref{algo:real-partical-protocol}.):
\begin{itemize}
\item In the \textit{first step},  we  construct an proxy procedure, named $\mathcal M'_Y(X, Y)$ as shown in Algorithm~\ref{algo:observe_card2}  with the same output probability mass function (PMF)  as that of $\mathcal M_Y(X, Y)$ in Algorithm~\ref{algo:real-partical-protocol}.
\item In the  \textit{second step}, we further reduce $\mathcal M'_Y(X, Y)$ in Algorithm~\ref{algo:observe_card2}, by replacing the computation involving $Y$ with a computation involving $I$, and prove the proxy $\mathcal M^{\prime\prime}_Y(X, Y)$ in Algorithm~\ref{algo:observe_card3}  assuring the same DP analysis for B as that in DP-PSI.

\item Finally, we demonstrate the $(\eB,\dB)$-DP assurance for the receiver in DP-PSI, as shown in Theorem~\ref{th:equal}, by considering all three parts of the randomness in Algorithm~\ref{algo:observe_card2}.

\end{itemize}

\noindent \textit{First Step}: In order to analyze the privacy for $Y$ in $\mathcal M_Y(X, Y)$, we  construct an equivalent procedure, named $\mathcal M^{\prime}_Y(X, Y)$ with the same output probability mass function (PMF)  as that of $\mathcal M_Y(X, Y)$.
Therefore, for any $z\in \mathbb Z$, $\mathcal M_Y(X, Y)$'s output  $\left|\mathsf I^{ab}_{\mathsf{sub}}\right|$  and $\mathcal M^{\prime}_Y(X, Y)$'s output $\mathsf z$ have the same PMF, i.e.,  $ \Pr\left[\left|\mathsf I^{ab}_{\mathsf{sub}}\right|=z\right]=\Pr\left[|\mathsf I^\prime_{\mathsf{sub}}|=z\right]$. 
Thus, $\mathcal M'_Y(X, Y)$ have the same DP guarantee. 
We summarize  $\mathcal M'_Y(X, Y)$ in Algorithm~\ref{algo:observe_card2}.
\begin{algorithm}
% \footnotesize
\SetKwInOut{Input}{Input}
\SetKwInOut{Output}{Output}
\Input{$X$, $Y$, $I=X\cap Y$.}
\Output{$|\mathsf I^\prime_{\mathsf{sub}}|$.}

% Sample $s \leftarrow \operatorname{Bin}\left(n, \frac{\lambda}{n}\right)$

{Sample $\mathsf s \leftarrow \operatorname{Bin}\left(n, 2(1-\pB)\right)$}
% should we write as Bernoulli trials??

Define   $\mathcal{T}_s = \{\mathsf{T}:\mathsf{T}\subseteq Y, |\mathsf{T}|=\mathsf s\}$

Choose $\mathsf T \leftarrow \mathcal{T}_{\mathsf s}$ uniformly at random

% Return $\Psi = \sum^n_{i\notin H,\hash(y_i)^{ab}\in X^{ab}_{\pi}}x_i+\operatorname{Bin}\left(s,\frac{1}{2}\right)$.\usepackage{algpseudocode}

% Return $\Psi = \sum_{y_i\in(B/T)\cap I} 1+\sum_{y_i\in T_1}\Ber\left(\frac{1}{2}\right)$

% Return $\mathsf z = | (B\setminus \mathsf T)\cap I| +\sum_{y_i\in \mathsf T\cap I}\Ber\left(\frac{1}{2}\right)$

Return {$|\mathsf I^\prime_{\mathsf{sub}}| = | I\setminus \mathsf T| +\sum_{y_i\in I\cap \mathsf T}\Ber\left(\frac{1}{2}\right)$}

\caption{$\mathcal M^{\prime}_{\mathsf{Y}}(X,Y)$, an equivalent proxy to   $\mathcal M_Y(X,Y)$ in  Algorithm~\ref{algo:real-partical-protocol} in terms of the output PMF.}
\label{algo:observe_card2}
\end{algorithm}

In Algorithm~\ref{algo:observe_card2} of $\mathcal M^{\prime}_Y(X, Y)$, a random set $\mathsf T \subseteq Y$ is first created.   The size of $\mathsf T$ is attributed to a Binomial distribution. $\mathsf T$ can be interpreted as the set that is endowed with the opportunity to do random response. To be more specific, if $y_i\in \mathsf T$ and $\hash(y_i)^{ab}\in X^{ab}_{\pi}$ (i.e., $y_i\in \mathsf T\cap  I$, where $I = X\cap Y$ is the ground-truth intersection between $X$ and $Y$), then it will be included to $I^{ab}_\mathsf{sub}$ with probability $\frac{1}{2}$. On the other hand, if $y_i\notin \mathsf T$ but $\hash(y_i)^{ab}\in X^{ab}_{\pi}$, it will be included to $I^{ab}_\mathsf{sub}$ honestly.
We formally  demonstrate in Lemma~\ref{claim:eq_card1_card2}
that in a probabilistic manner of the output, 
 Algorithm~\ref{algo:observe_card2} is equivalent to Algorithm~\ref{algo:real-partical-protocol}. Details are provided in Appendix~\ref{app:1st-equ}.

\begin{lemma}\label{claim:eq_card1_card2} The PMF  of $|\mathsf I^{ab}_{\mathsf{sub}}|$ and $|\mathsf I^\prime_{\mathsf{sub}}|$, which are the outputs of $\mathcal M_Y(X, Y)$ in Algorithm~\ref{algo:real-partical-protocol} and $\mathcal M^{\prime}_Y(X, Y)$ in Algorithm~\ref{algo:observe_card2},  are equal, i.e.,
$$
    \Pr\left[\left|\mathsf I^{ab}_{\mathsf{sub}}\right|=z\right]=\Pr\left[|\mathsf I^\prime_{\mathsf{sub}}|=z\right],  \quad \forall z\in \mathbb{Z}_{\geq 0}.
$$ 
\end{lemma}

\noindent \textit{Second Step}:
To analyze the privacy for $Y$ in Algorithm~\ref{algo:observe_card2}, we further define $\mathsf T_1$, a random subset of $I$ ($I\triangleq X\cap Y$), by $\mathsf T_1\triangleq \mathsf T\cap I$ and meanwhile $\mathsf T_2 \triangleq \mathsf T\setminus \mathsf T_1$. It is then evident that $|\mathsf T_1|+|\mathsf T_2|=|\mathsf T|\triangleq \mathsf s$. We also define the associated cardinalities by $\mathsf s_1 \triangleq |T_1|$ and $\mathsf s_2 \triangleq |T_2|$. Since the sum of Bernoulli random variables is  a Binomial random variable and  
$\mathsf s = \operatorname{Bin}(n, 2(1-p))$ in Algorithm~Algorithm~\ref{algo:observe_card2}, we have $\mathsf s_1 = \operatorname{Bin}(|I|, 2(1-p))$ and 
$\mathsf s_2 = \operatorname{Bin}(|B\setminus I|, 2(1-p))$.
We, therefore, represent $\mathsf T$ in Line 3 of Algorithm~\ref{algo:observe_card2} by two sets $\mathsf T_1$ and $\mathsf T_2$.
Moreover, the output of  Algorithm~\ref{algo:observe_card2} (Line 5) can be further simplified by
$|\mathsf I^\prime_{\mathsf{sub}}| = | I\setminus \mathsf T_1| +\sum_{y_i\in I\cap \mathsf T_1}\Ber\left({1}/{2}\right)$.
Consequently, instead of considering  all the the receiver's data, i.e., $\mathsf T\subseteq Y$, we can focus on only  the part in $I$, i.e., $\mathsf T_1\subseteq I$, and arrive at a more compact represent in Algorithm~\ref{algo:observe_card3}, which will be proved to be    equivalent to $\mathcal M^{\prime}_Y(X, Y)$ (Algorithm~\ref{algo:observe_card2}) in Lemma~\ref{claim:eq_card2_card3}.
The proof is in Appendix~\ref{app:2nd-equ}.

\begin{algorithm}
% \footnotesize
\SetKwInOut{Input}{Input}
\SetKwInOut{Output}{Output}
\Input {$X$, $Y$, $I=X\cap Y$, $\mathsf T_1=\emptyset$}
\Output{$|\mathsf I^{\prime\prime}_{\mathsf{sub}}|$.}

% Sample $s \leftarrow \operatorname{Bin}\left(I, \frac{\lambda}{n}\right)$

{Sample $\mathsf s_1 \leftarrow \operatorname{Bin}\left(|I|, 2(1-\pB)\right)$}

% \ForAll{$y_i\in I$}{

% Let $\mathsf{x}_i \leftarrow \Ber\left(2(1-p)\right)$

% \If{$\mathsf x_i=1$}{
% $\mathsf y_i=1$ and $\mathsf T_1 = \mathsf T_1\cup y_i$
% }
% \Else {$\mathsf y_i=0$}
% }

% $ \mathsf s_1=\sum_{y_i\in I}\mathsf y_i$

{Let $\mathcal{T}_{s_1} = \{\mathsf T_1: \mathsf T_1\subseteq I,|\mathsf T_1|=s_1\}$}

Choose $\mathsf T_1 \leftarrow \mathcal{T}_{s_1}$ uniformly at random

% Return $\Psi = \sum^n_{i\notin H,\hash(y_i)^{ab}\in X^{ab}_{\pi}}x_i+\operatorname{Bin}\left(s,\frac{1}{2}\right)$.\usepackage{algpseudocode}

% Return $\Psi = \sum_{y_i\in(B/T)\cap I} 1+\sum_{y_i\in T_1}\Ber\left(\frac{1}{2}\right)$

{Return $
|\mathsf I^{\prime\prime}_{\mathsf{sub}}|= | I\setminus \mathsf T_1| +\sum_{y_i\in \mathsf T_1}\Ber\left({1}/{2}\right)$}

% \hl{Return $\mathsf z = | (B\setminus T)\cap I| +\sum_{y_i\in T_1}\Ber\left(\frac{1}{2}\right)$}

% Return $\Psi = \sum^n_{i\notin H,\hash(y_i)^{ab}\in X^{ab}_{\pi}}x_i+\sum^n_{i\in H,\hash(y_i)^{ab}\in X^{ab}_{\pi}}\Ber\left(\frac{1}{2}\right)$.

\caption{$\mathcal M^{\prime\prime}_{\mathsf{Y}}(X, Y)$: an equivalent to Algorithm~\ref{algo:observe_card2}}
\label{algo:observe_card3}
\end{algorithm}

\begin{lemma}\label{claim:eq_card2_card3} 
The PMF  of $|\mathsf I^{\prime}_{\mathsf{sub}}|$ and $|\mathsf I^{\prime\prime}_{\mathsf{sub}}|$ 
, which are the outputs of $\mathcal M^{\prime}_Y(X, Y)$ in Algorithm~\ref{algo:observe_card2} and $\mathcal M^{\prime\prime}_Y(X, Y)$ in Algorithm~\ref{algo:observe_card3}, respectively, are equal, i.e.,
\begin{equation*}
    \Pr[|\mathsf I^\prime_{\mathsf{sub}}|=z]=\Pr[|\mathsf I^{\prime\prime}_{\mathsf{sub}}|=z],   \quad \forall z\in \mathbb{Z}_{\geq 0}.
\end{equation*}
% \begin{equation*}
% \begin{aligned}
%     &\kappa\Pr[\mathsf{z}_1=z] = \Pr[\mathsf z=z],\\
%     \textit{where} & \quad \kappa  = \left(1-2(1-p)\right)^{n-|I|}  \left(1+\frac{2(1-p)}{1-2(1-p)}\right)^{|B\setminus I|}= 1.
% \end{aligned}
% \end{equation*}
\end{lemma}

According to the DP definition, Lemma~\ref{claim:eq_card1_card2}  and Lemma~\ref{claim:eq_card2_card3} demonstrate that 
$|\mathsf I^{ab}_{\mathsf{sub}}|$, $|\mathsf I^\prime_{\mathsf{sub}}|$, and 
$|\mathsf I^{\prime\prime}_{\mathsf{sub}}|$ all share the same privacy guarantee for the receiver, , and we draw this conclusion formally in the following theorem.

\begin{theorem}
\label{th:equal}
The output  $|\mathsf I^{ab}_{\mathsf{sub}}|$ in Algorithm~\ref{algo:real-partical-protocol} and the output  $|\mathsf I^{\prime\prime}_{\mathsf{sub}}|$ in Algorithm~\ref{algo:observe_card3} are  identically distributed, i.e., they have the same PMF. Algorithm~\ref{algo:real-partical-protocol} and Algorithm~\ref{algo:observe_card3}, therefore, ensure the same $(\eB,\dB)$-DP for $Y=[b_1,b_2,\ldots,b_n]$.
\end{theorem}

\noindent \textit{Third Step}:
We emphasis that $\mathcal M^{\prime}_Y(X, Y)$ (Algorithm~\ref{algo:observe_card2}) and 
$\mathcal M^{\prime\prime}_Y(X, Y)$ (Algorithm~\ref{algo:observe_card3}) are both  proxy procedures for DP analysis,
which are not executed during the implementation ofDP-PSI. 
InDP-PSI, the randomness of $|I^{ab}_{\mathsf{sub}}|$ is entirely due to the Bernoulli subsampling of $Y$, whereas, in Algorithm~\ref{algo:observe_card3}, such randomness is cast into three parts summarized as below: 

\noindent \textit{(a)  Sum of Bernoulli r.v., i.e., $\sum_{y_i\in T_1}\Ber\left({1}/{2}\right)$  with  $s_1$ and $T_1$ held  constant.}

\noindent \textit{(b) Sum of Bernoulli r.v. and randomly chosen $\mathsf T_1$, yet with a fixed  $s_1$.}

\noindent \textit{(c)  Sum of Bernoulli r.v., randomly chosen $\mathsf T_1$ and randomly sampled $\mathsf s_1$ by $\mathsf s_1 \leftarrow \operatorname{Bin}\left(|I|, 2(1-\pB)\right)$.}

In what follows, we provide the DP analysis of $\mathcal M^{\prime\prime}_{\mathsf{Y}}(X, Y)$ by examining the privacy guarantee of the composition of all the above randomness.

\noindent\textit{(a) Sum of Bernoulli r.v.} To facilitate the DP analysis, we tease out Line~4 from Algorithm~\ref{algo:observe_card3} with a fixed $T_1\in I$ as detailed in Algorithm~\ref{algo:decomp1}.
The output $\mathsf z_1$, as specified in the following Lemma~\ref{claim:C_T_dp}, is $(\epsilon_{\mathsf{z}_{1}},\frac{\delta}{2})$-DP guarantee with proof provided in Appendix~\ref{app:1st-random}.
% It is interesting to note that 
% and  achieve Lemma~\ref{claim:C_T_dp}.

\begin{algorithm}
% \footnotesize
\SetKwInOut{Input}{Input}
\SetKwInOut{Output}{Output}
\Input {Fix $T_1$ in Line~3 of Algorithm~\ref{algo:observe_card3} with $T_1 \subseteq I$ 
}
\Output{$\mathsf{z}_{1}$}

% Let $\mathbf{B}$ $\leftarrow \operatorname{Bin}\left(|H|,\frac{1}{2}\right)$

% \textbf{Return} $\mathbf{y}_T = \sum^n_{i\notin H,\hash(y_i)^{ab}\in X^{ab}_{\pi}}x_i+\mathbf{B}$.

% Return $\Psi = \sum^n_{i\notin H,\hash(y_i)^{ab}\in X^{ab}_{\pi}}x_i+\sum^n_{i\in H,\hash(y_i)^{ab}\in X^{ab}_{\pi}}\Ber\left(\frac{1}{2}\right)$.

% Return $\Psi = \sum_{y_i\in(B/T)\cap I}1+\sum_{y_i\in T_1}\Ber\left(\frac{1}{2}\right)$.

Return $\mathsf{z}_{1} = |I\setminus T_1|+{\sum_{y_i\in T_1}\Ber\left(\frac{1}{2}\right)}$

\caption{Fixed $T_1$ in  Algorithm~\ref{algo:observe_card3}% \hl{(decomposition of algo 2, the first component, i.e., considering the randomness of $|H|$, $H$ can be interpreted as the set of individuals who get the opportunity to do RR)}
}
\label{algo:decomp1}
\end{algorithm}

\begin{lemma}\label{claim:C_T_dp}
For any fixed set $T_1$ with $T_1\subseteq I$ and $\delta>0$,  %such that 
% {\color{blue}\begin{equation*}
%     |T_1|\geq \max\left\{\frac{1}{2}\left(\log\frac{4}{\delta}+\sqrt{\log\frac{4}{\delta}+4}\right)^2, 18\log \frac{4}{\delta}\right\},
% \end{equation*}}
% {\color{blue}\begin{equation*}
%     |T_1|\geq 18\log \frac{4}{\delta},
% \end{equation*}}
Algorithm~\ref{algo:decomp1} is $(\epsilon_{\mathsf{z}_{1}},\frac{\delta}{2})$-DP for 
\begin{equation}
\epsilon_{\mathsf{z}_{1}} = \frac{\frac{|T_1|}{2}+\sqrt{\frac{|T_1|}{2}\log\frac{4}{\delta}}}{\frac{|T_1|}{2}-\sqrt{\frac{|T_1|}{2}\log\frac{4}{\delta}}}.
\end{equation}
\end{lemma}

\noindent\textit{(b) Sum of Bernoulli r.v. and randomly chosen $\mathsf T_1$, yet with a fixed  $s_1$.} We incorporate the second randomness, where $\mathsf T_1$ is a randomly selected subset of $I$ in Algorithm~\ref{algo:observe_card3}. In Lemma~\ref{claim:dp_amp_sub}, we conclude that this randomness improves the privacy in Lemma~\ref{claim:C_T_dp} by a factor of $1-2(1-p)$ due to privacy amplification \cite{raskhodnikova2008can}. In Algorithm~\ref{algo:decomp2}, we summarize the steps concerning the subset $T_1$, and complete the proof in Appendix~\ref{app:2nd-random}.

\begin{algorithm}
% \footnotesize
\SetKwInOut{Input}{Input}
\SetKwInOut{Output}{Output}
\Input {A constant $s_1$ in Line 2 of Algorithm~\ref{algo:observe_card3}}
\Output{$\mathsf z_{2}$.}

% Sample $s_1 \leftarrow \operatorname{Bin}\left(|I|, 2(1-p)\right)$

Define   $\mathcal{T} = \{\mathsf T_1\subseteq I:|\mathsf T_1|=s_1\}$

Choose $\mathsf T_1 \leftarrow \mathcal{T}$ uniformly at random

% Return $\Psi = \sum^n_{i\notin H,\hash(y_i)^{ab}\in X^{ab}_{\pi}}x_i+\operatorname{Bin}\left(s,\frac{1}{2}\right)$.\usepackage{algpseudocode}

% Return $\Psi = \sum_{y_i\in(B/T)\cap I} 1+\sum_{y_i\in T_1}\Ber\left(\frac{1}{2}\right)$

Return $\mathsf z_{2} = | I\setminus {\mathsf  T_1}| +{\sum_{y_i\in \mathsf T_1}\Ber\left(\frac{1}{2}\right)}$

\caption{Fixed $s_1$ in  Algorithm~\ref{algo:observe_card3}}
\label{algo:decomp2}
\end{algorithm}

% \begin{equation}
% \mathcal S_a\triangleq
% \left\{\mathsf s_1: \mathsf s_1>\max\left\{2+\log\frac{4}{\delta}+\sqrt{4\log\frac{4}{\delta}+\left(\log\frac{4}{\delta}\right)^2},18\log\frac{4}{\delta}\right\}\right\}.
% \end{equation}
\begin{lemma}\label{claim:dp_amp_sub}
For any $\delta>0$, %and $s_1\in \mathcal S_a$, 
the output $\mathsf z_{2}$ in Algorithm~\ref{algo:decomp2}
 is $(\epsilon_{\mathsf z_{2}},\frac{\delta}{2})$-DP for $$\epsilon_{\mathsf z_{2}} =  \frac{\frac{s_1}{2}+\sqrt{\frac{s_1}{2}\log\frac{4}{\delta}}+1}{\frac{s_1}{2}-\sqrt{\frac{s_1}{2}\log\frac{4}{\delta}}} - 1.$$

% and $\epsilon_8 = \frac{\frac{|T_1|}{2}-\sqrt{\frac{|T_1|}{2}\log\frac{4}{\delta}}}{\frac{|T_1|}{2}+\sqrt{\frac{|T_1|}{2}\log\frac{4}{\delta}}}$%$\epsilon = \left(1-2(1-p)\right) \sqrt{\frac{18\log\frac{4}{\delta}}{s_1}}$.
\end{lemma}

In contrast to the case in Lemma~\ref{claim:dp_amp_sub}, the extra randomness of selecting $\mathsf T_1$ ensured DP amplification. Specifically, when $T_1=s 1$, such randomness reduces the $\epsilon$ by one.

\noindent\textit{(c) sum of Bernoulli r.v., randomly chosen $\mathsf T_1$, and randomly sampled $\mathsf s_1$ by $\mathsf s_1 \leftarrow \operatorname{Bin}\left(|I|, 2(1-\pB)\right)$.} 
We discuss the privacy guarantee of $\mathcal M^{\prime\prime}_{\mathsf{Y}}(X, Y)$ in Algorithm~\ref{algo:observe_card3}, where the randomness includes Moreover, according to 
Theorem~\ref{th:equal}, we have the equivalent $(\eB,\dB)$-DP for the receiver in DP-PSI that is summarized in  Algorithm~\ref{algo:real-partical-protocol}. 
% The formal claim is provided in the lemma below with  analysis details shown in Appendix~\ref{app:3rd}.

\begin{theorem}
\label{claim:dp_I_int_AB}
For any $\dB>0$ and $\pB\in(0,1)$, 
DP-PSI (equivalently $\mathcal M_{\mathsf{Y}}(X, Y)$ in Algorithm~\ref{algo:real-partical-protocol}
and
$\mathcal M^{\prime\prime}_{\mathsf{Y}}(X, Y)$ in Algorithm~\ref{algo:observe_card3}) has  $(\eB,\dB)$-DP guarantee for the receiver with
\begin{equation}
\label{eq:eY_range}
\eB = \frac{2\sqrt{t\log\frac{4}{\dB}}+1}{t-\sqrt{t\log\frac{4}{\dB}}},
\end{equation}
where
$t \triangleq (1-\pB)|I|-\sqrt{\frac{|I|}{8}\log\frac{2}{\dB}}$, if $|I|\geq |I_L|$. The  cardinality lower bound for $I$ is denoted by $|I_L|$ and defined by
\begin{equation}
\label{eq:IL}
|I_L|>
\frac{\left(\sqrt{\frac{1}{2}\log\frac{2}{\dB}} + 
\sqrt{\frac{1}{2}\log\frac{2}{\dB}+16(1-p)\log\frac{4}{\dB}}\right)^2 }
{16(1-\pB)^2}.
\end{equation}
\end{theorem}

% \begin{IEEEproof}
% Let   $\Pr[\mathsf{z}_1=z]$ be the PMF of the output of $\mathcal M^{\prime\prime}_{\mathsf{Y}}(X, Y)$ in Algorithm~\ref{algo:observe_card3}, $\Pr[\mathsf{z}=z]$ be the PMF of the output of $\mathcal M^{\prime}_{\mathsf{int}}(X, Y)$ in Algorithm~\ref{algo:observe_card2}, and $\Pr[|\mathsf I^{ab}_{\mathsf{sub}}|=z]$ be the PMF  of $|\mathsf I^{ab}_{\mathsf{sub}}|$ taking value $z$ in $\mathcal M_{\mathsf{int}}(X, Y)$ in Algorithm~\ref{algo:real-partical-protocol}. Then   we have
% \begin{equation*}
%   \frac{\Pr[\psi(B)=z]}{\Pr[\psi(\widetilde{B})=z]} \stackrel{(a)} =  \frac{\Pr[\mathsf{z}(B)=z]}{\Pr[\mathsf{z}(\widetilde{B})=z]} \stackrel{(b)}=  \frac{ \Pr[\mathsf{z_1}(B)=z]}{ \Pr[\mathsf{z_1}(\widetilde{B})=z]}  \stackrel{(b)}\leq e^\epsilon,
% \end{equation*}
% where $(a)$,  $(b)$ and $(c)$ are due to Claim \ref{claim:eq_card1_card2}, 
%  Claim \ref{claim:eq_card2_card3}, and Cliam \ref{claim:dp_I_int_AB},  respectively. 
% \end{IEEEproof}

% \clearpage

% \clearpage
\subsection{$\mathcal M_{\mathsf X}$, Privacy for  $X$}
\begin{algorithm}
% \footnotesize
\SetKwInOut{Input}{Input}
\SetKwInOut{Output}{Output}
\Input {$X$, $Y$, $\hash(\cdot)^x$, $\pA$, $\pB$, $q$, $ a\sample\ZZ_p$, and $b\sample\ZZ_p$ .} %Y^{ab}_\mathsf{sub}$
\Output{$\mathsf I_{\mathsf{DP}}$.}

The receiver obtains $\bar {\mathsf Y}_{\mathsf sub}$ by first re-order $Y$ to obtain $Y$ and then subsampling  $Y$ by $\mathsf{Ber}(Y; \pB)$.

The sender computes the \textbf{intersection}: $  {\mathsf I^{ab}_\mathsf{sub}}\triangleq X^{ab}_{\pi}\cap {\mathsf{Y}^{ab}_\mathsf{sub}}$.

The sender computes the \textbf{sub-sampling}: $\widetilde{\mathsf I}_\mathsf{sub}^{ab}\gets\Ber(\mathsf I_\mathsf{sub}^{ab};\pA)$.
 
The sender computes the \textbf{up-sampling} as $\mathsf I_\mathsf{DP}^{ab}\gets\Ber(\mathsf{Y}_\mathsf{sub} \setminus \mathsf I^{ab}_\mathsf{sub}; q)\cup \widetilde{\mathsf I}_\mathsf{sub}^{ab}$  and sends to the receiver the  elements' indices of ${\mathsf I_{\mathsf {DP}}^{ab}}$ in 
         $\mathsf{Y}_\mathsf{sub}^{ab}$  with lexicographical order.

The receiver obtains $\mathsf I_{\mathsf{DP}}$

\caption{$\mathsf I_{\mathsf{DP}}\triangleq \mathcal M_{\mathsf X}(X, Y)$: DP for the sender in $\mathrm{DP^2SI}$ output.} 
\label{algo:real-DPA}
\end{algorithm}
Following the DP-PSI protocol, it is guaranteed that for any $x_i\in X$ and $x_i\notin X\cap Y_{\mathsf{sub}}$,  we have
$x_i\notin I_{\mathsf{DP}}$. The receiver, therefore, gets no information of any $x_i$ with  $x_i\notin I_{\mathsf{DP}}$.
And the privacy for such  $x_i\notin X\cap Y$ is  perfectly guaranteed with the view of $\IDP$. We therefore focus on the privacy protection for $x_i\in X\cap Y_{\mathsf{sub}}$.
For any $y_i\in Y_{\mathsf{sub}}$, we define the  r.v. $X_i$ as follows.
\begin{itemize}
    \item $X_i = 1$ denotes the case that $y_i\in \IDP$.  Equivalently, there exists an $a_\ell$ in the DP-PSI output such that $a_\ell = y_i$.
    \item
    $X_i = 0$ denote the case that $y_i\notin \IDP$. Equivalently, there does not exist any $a_\ell\in X$ such that $a_\ell = y_i$.
\end{itemize}
Following the DP-PSI protocol,  there are four possible $X_i$ given the corresponding $y_i\in Y_{\mathsf{sub}}$:

% \begin{itemize}
% \begin{enumerate}
\noindent 1) 
There exists an $a_{\ell}\in X$ such that $a_{\ell} = y_i$. This is the case that corresponds to the truth positive and we represent it in terms of the conditional probability by
\begin{equation}
\label{eq:TP}
\Pr(X_i=1|\exists a_{\ell}=y_i)=\pA.
\end{equation}

\noindent 2)  There exists an $a_{\ell}\in X$ such that $a_{\ell} = y_i$ but $X_i=0$. This is the case that corresponds to the false negative and we represent it in terms of the conditional probability by
\begin{equation}
\label{FN}
\Pr(X_i=0|\exists a_{\ell}=y_i)=1-\pA.
\end{equation}

\noindent 3)  There is no element  $a_{\ell}\in X$ such that  $a_{\ell} = y_i$ and $X_i=0$. Such $X_i=0$ must come from the up-sampling  in the DP-PSI protocol, which corresponding to the truth negative, we have
\begin{equation}
\label{eq:TN}
\begin{split}
\Pr(X_i=0|\nexists  a_{\ell}=y_i)
=1-q.
\end{split}
\end{equation}

% \begin{equation}
% \label{eq:FP}
% \Pr\left(X_{i}=1 \mid {\mathbbm 1}_{X\cap  Y_{sub}}(a_{\ell})=0\right)=q.
% \end{equation}

\noindent 4)  There is no element  $a_{\ell}\in X$ such that  $a_{\ell} = y_i$ and $X_i=1$. Such $X_i=1$ must come from the up-sampling  in the DP-PSI protocol, which corresponding to the false positive, we have
\begin{equation}
\label{FP}
\Pr(X_i=1|\nexists  a_{\ell}=y_i)
= q.
\end{equation}
Given the above four possible outcomes, the privacy for the sender in DP-PSI follows the classical randomized response mechanism, and we provide the $\eA$-DP given the valid region as shown in the following theorem with details in Appendix~\ref{app:rr}.

\begin{theorem}
\label{lemma:region}
Given $\bar Y_{\mathsf{sub}}$ at the receiver, the  DP-PSI protocol satisfies $\eA$-DP for $X=[x_1,\ldots,x_n]$ if $\pA$ and $q$ belongs to the region $\mathcal R$ as follows:
\begin{equation}
\label{eq:epsilon-region}
\mathcal R = 
\left\{
             \begin{array}{lr}
\pA\leq e^{\eA}q & \\ 
1-q  \leq e^{\eA}\pA &\\ 
0\leq \pA, q \leq 1 &
             \end{array}.
\right.
\end{equation}
\end{theorem}

\begin{proposition}
\label{prop:increasing}
The function $t = (1-\pB)|I|-\sqrt{\frac{|I|}{8}\log\frac{2}{\dB}}$ is 
a monotonically increasing function w.r.t. $|I|$ for 
$|I|>\frac{1}{8(1-\pB)^2}\log\frac{2}{\dB}.$   
\end{proposition}

\begin{proposition}
\label{prop:decreasing}
The function $\eB$ in (\ref{eq:eY_range}) is monotonically decreasing w.r.t. $t$ for $t>\log \frac{4}{\dB}$.
\end{proposition}

According to Propostion~\ref{prop:increasing} and Propostion~\ref{prop:decreasing}, $\eB$ is monotonic decreasing function of $|I|$.
Therefore, the smaller the $|I|$, the larger $\eB$.
We, therefore, immediately have the  upper bound of $\eB$ given by the following lemma.
\begin{lemma}
For any $1>\dB>0$, when 
\begin{equation}
|I|>
\frac{\left(\sqrt{\frac{1}{2}\log\frac{2}{\dB}} + 
\sqrt{\frac{1}{2}\log\frac{2}{\dB}+16(1-\pB)\log\frac{4}{\dB}}\right)^2 }
{16(1-\pB)^2},
\end{equation}  
DP-PSI achieves $(\eB, \dB)$-DP with  
\begin{equation}
\label{eq:eb2}
\eB = \frac{2\sqrt{t\log\frac{4}{\dB}}+1}{t-\sqrt{t\log\frac{4}{\dB}}},
\end{equation}
where 
$t = (1-\pB)|I_L|-\sqrt{\frac{|I_L|}{8}\log\frac{2}{\dB}}$, and  $|I_L|$ is defined in Eq (\ref{eq:IL}).
\end{lemma}

\noindent 2) Choosing $\pA$ and $q$

According to Theorem~\ref{lemma:region}, there are an infinite number of ($\pA$, $q$) pairs that can achieve the same $\eA$-DP.
To achieve optimal utility given a fixed $\eA$, we make an chosen of $\pA$ and $q$ aiming to achieve better PSI result of $\IDP$ in terms of precision and recall.
Specifically, precision defines the percentage of elements in $\IDP$ come from  $X\cap Y_{\mathsf{sub}}$, which is given by
\begin{eqnarray}
\label{eq:precision}
% \begin{split}
\text{precision}&\triangleq
&\frac{|\Ber(X^{ab}_{\pi}\cap {\mathsf{Y}^{ab}_\mathsf{sub}};\pA)| }
{
|\Ber(\mathsf{Y}^{ab}_\mathsf{sub} \setminus \mathsf I^{ab}_\mathsf{sub}; q)| + |X^{ab}_{\pi}\cap {\mathsf{Y}^{ab}_\mathsf{sub}}|}\nonumber\\
&\approx&
\frac{\pA|X\cap Y_{\mathsf{sub}}| }
{
q|Y_{\mathsf{sub}}\setminus X| + \pA| X\cap Y_{\mathsf{sub}}|}.
% \end{split}.
\end{eqnarray}
The approximation results from computing the expectations of numerator and denominator  separately.
Meanwhile, 
recall defines
 what percentage of elements in $X\cap Y_{\mathsf{sub}}$ is finally in $\IDP$, which is given by
\begin{eqnarray}
\label{eq:recall}
% \begin{split} 
\text{recall}
&\triangleq&
\frac{|X^{ab}_{\pi}\cap {\mathsf{Y}^{ab}_\mathsf{sub}}| }
{
q|\bar{\mathsf Y}^{ab}_{\mathsf{sub}}\setminus X^{ab}_{\pi}| + |X^{ab}_{\pi}\cap {\mathsf{Y}^{ab}_\mathsf{sub}}|}\nonumber\\
&\approx&
\frac{\pA|X\cap \mathsf Y_{\mathsf{sub}}| }
{
q|\mathsf Y_{\mathsf{sub}}\setminus X| + \pA| X\cap \mathsf Y_{\mathsf{sub}}|},
% \end{split}
\end{eqnarray}
where the approximation is due to computing the numerator and denominator expectations separately.
The following lemma gives the specific values for $\pA$ and $q$ by restricting $\pA$ and $q$ in region $\mathcal R$ in Eq.~(\ref{eq:epsilon-region}). The computation of optimal $\pA$ and $q$ is similar to that in \cite{}.
We provide the details in Appendix~\ref{app:precision-recall}.

\begin{lemma}
\label{optimal_p_q}
Given a $\mathsf Y_{\mathsf{sub}}$, the  precision and the corresponding recall in Eq.~(\ref{eq:precision}) and Eq.~(\ref{eq:precision}) for the DP set intersection $\mathsf I_{\mathsf{DP}}$ are
\begin{equation}
\label{precision-recall}
    \textrm{precision}^{\ast} =
    \frac{|\mathsf I_{\mathsf {sub}}| }
{
e^{-\eA}|\mathsf Y_{\mathsf{sub}}\setminus X| + |\mathsf I_{\mathsf {sub}}|},
 \quad
    \textrm{recall}^{\ast} = 
    \frac{e^{\eA}}{1+e^{\eA}}.
\end{equation}
with the optimal $\pA$ and $q$ given by:
\begin{equation}
\label{optimal_p}
\pA^\ast = \frac{e^{\eA}}{1+e^{\eA}},
\quad \textrm{and} \quad
q^\ast =\frac{1}{1+e^{\eA}}.
\end{equation}
\end{lemma}

\section{EVALUATIONS}
\label{sec:8}
% \subsection{PSI}
In this experiment we implemented our DP-PSI method between the sender and the receiver. We simulated these two parties in a server with Intel(R) Xeon(R) Platinum 8269CY CPU T 3.10GHz 64 cores and 512GB memory. 
We tested our protocol between two parties with equal data set. The number of elements in each data set increased from $2^{10}$ to $2^{28}$, which proves the practicality of our solution for very large-scale implementations. The ratio of matching element between two parties was about 70\%.
The elliptic curve we chose for our PSI protocol is curve25519. All the parameters for our ECC implementation can be found at TLS Standard.

The detail of our experimental results is presented in Table 1 and Table 2.   From the results we can find that the communication cost grows linearly with the increase of the element number. Since we implemented our ECC PSI with curve25519, the running time also roughly grows linearly with the increasing rate of element number. Since the running time to shuffle data elements, sub-sampling and up-sampling can be neglected, the results show that our protocol is applicable for adding DP protection to data set with elements up to $2^{27}$ (about 134 million). We believe this scale of data is already large enough for most real-world DP-PSI application. On contract, the solution proposed in \cite{kacsmar2020differentially} was only applicable for data set with less than $2^16$ elements on server’s side due to the expensive operation. On the client side, the number of elements was even less.

\begin{table*}[htb!]
\centering
\begin{adjustbox}{width=0.6\textwidth}
\hfill{}\begin{tabular}{|c|c|c|c|c|}
\hline 
\multirow{2}{*}{} & \multicolumn{2}{c|}{Kacsmar et al.'\cite{kacsmar2020differentially}}  & \multicolumn{2}{c|}{{DP-PSI} ($\epsilon = 3$)}
\tabularnewline
\cline{2-5} 
Input Size: $n=2^k$ & {Runtime (sec)} &  {Comm.(MB)}  & {Runtime (s)} & {Comm.(MB)} 
\tabularnewline
\hline 
{$k=10$} &  84.46   & 111.66      & 0.012  & 0.074   \tabularnewline
\hline 
{$k={11}$} &  160.07  &  169.02    &0.022 &0.152   \tabularnewline
\hline 
{$k={12}$} & 413.72  & 289.35      &0.041  &0.305 \tabularnewline
\hline 
{$k={13}$} &  740.47  &  541.84      & 0.073  &   0.607 \tabularnewline
\hline 
{$k={14}$} &  1537.83  & 1083.73  & 0.141  & 1.21   \tabularnewline
\hline 
{$k={15}$} & 3610.24  & 2218.59       & 0.282 &2.43  
\tabularnewline
\hline 
{$k={16}$} & -  &   -     & 0.565  &4.85  
\tabularnewline
\hline 
{$k={17}$} & -  &   -     & 1.136  &9.71  
\tabularnewline
\hline
\end{tabular}\hfill{}
\end{adjustbox}
\caption{Performance comparison with the method without payload computation.}
\label{table:Ion&Pinkas}
\end{table*}

% Setting 1: $\epsilon_2=3$,
% $\epsilon_3 = 3$, $\delta_1=\frac{1}{10n}$ and compute 

% Setting 2: $\epsilon_2=4$
% $\epsilon_3 = 3$, and $\delta_1=\frac{1}{20n}$
% $\delta_3=\frac{1}{20n}$
% \subsection{Statistics}

\begin{table*}[htb!]
\centering
\begin{adjustbox}{width=0.8\textwidth}
\hfill{}\begin{tabular}{|c|c|c|c|c|c|c|c|c|c|}
\hline 
\multirow{2}{*}{} & \multicolumn{2}{c|}{Ion et al.'\cite{ion2020deploying}} & \multicolumn{2}{c|}{Pinkas et al.' \cite{pinkas2019efficient}} & \multicolumn{2}{c|}{{DP-PSI} ($\epsilon = 3$)}
\tabularnewline
\cline{1-2} \cline{2-4} \cline{5-7} \cline{8-10}
Input Size: $n=2^k$ & {Time (sec)} &  {Comm.(MB)} & {Time (sec)} & {Comm.(MB)} & {Time (sec)} & {Comm.(MB)} 
\tabularnewline
\hline 
{$2^{18}$} & 29.9   & 84.8   &4.5   &113.5    & 2.273  & 19.43   \tabularnewline
\hline 
{$2^{19}$} & 59.5  &169.5 & 8.6 & 227.6  & 4.573 &38.85   \tabularnewline
\hline 
{$2^{20}$} & 120.2 & 339.0  & 14.7 &456.6   &9.801  &81.27 \tabularnewline
\hline 
{$2^{21}$} &  242.3 & 678.1   &28.8   &915.8    & 20.461  &   162.54 \tabularnewline
\hline 
{$2^{22}$} & 484.7  & 1356.1& 54.8 &1836.9   & 42.234&325.07   \tabularnewline
\hline 
{$2^{23}$} & 965.6 & 2712.24  &112.6  &3684.5   &86.625  &625.15 \tabularnewline
\hline 
{$2^{24}$} & - &  - &235.9  &7390.5   &176.384  & 1300.27\tabularnewline
\hline 
{$2^{25}$} & - & -  &464.3  & 14824  &345.647  & 2600.51\tabularnewline
\hline 
{$2^{26}$} & - & -  &939.7  &29734   &724.819 &5201.05 \tabularnewline
\hline 
{$2^{27}$} & - &  - & - & - & {1475.652}  &10402.06 \tabularnewline
\hline 
\end{tabular}\hfill{}
\end{adjustbox}
\caption{Performance comparison with PSI methods with payload computation.}
\label{table:Ion&Pinkas}
\end{table*}

The second experiment we did was to compare out solution with~\cite{ion2020deploying} and~\cite{pinkas2019efficient}, which is shown in Table~\ref{table:Ion&Pinkas}. Not only we implemented their PSI solutions in the same environment, but also we did a sum on the payload associated with the matching elements with DP protection on both matching elements and payloads. Google in~\cite{ion2020deploying} named this PSI with payload sum as Private Intersection-Sum (PIS) and argued that PIS is very important for Ads privacy, logins/passwords privacy, etc.~\cite{google-psi}.  Here we highlight the difference between our solution and theirs. They either used full shuffle ~\cite{ion2020deploying}  or MPC circuit ~\cite{pinkas2019efficient} to secretly hide the intersection and only compute the sum of payloads associated with the matching elements either with partial HE algorithm ~\cite{ion2020deploying}  or MPC circuits ~\cite{pinkas2019efficient}. While our solution achieves DP protection to intersection and then discloses the DP protected matching element to one party and then directly computing the corresponding sum of the payloads in plaintexts. It is clear that our solution DOES NOT protect the intersection with a cryptographic approach. However, we argue that our approach is still useful for scenarios where the receiver needs the intersection results such as in advertising  like mentioned in~\cite{google-psi}. That said, adding DP protection to the intersection might be a must. In this case, we provide an alternative to  protecting both intersection and the associated payloads via  DP. Table~\ref{table:Ion&Pinkas} shows the results. We combined the running time of PSI and computing the sum of payloads and our solution is not only scalable to large-scale data sets but also much more effective than  other approaches. 
\newpage
% \newpage
% \input{sections/8_exp}
\appendix
% \subsection{An example of the D$\text P^2$SI protocol}
% In Fig.~\ref{fig:dp-psi-example}....

% \begin{figure*}[htbp]
% \centering
% \includegraphics[width=0.7\textwidth]{images/DPSI-example.jpg}
% \caption{An example of  the D$\text P^2$SI protocol with
% Alice's input  $A= [a, b, c, d, e, h]$,  Bob's input  $B= [b, c, d, f, g, h]$, and the the D$\text P^2$SI output result $[d, f, h]$.
% }
% \label{fig:dp-psi-example}
% \end{figure*}

\subsection{Proof of Lemma~\ref{lemma:security}}
\label{app:lemma:security}

Recall the lemma from the original DH-PSI paper \cite{DBLP:conf/sigmod/AgrawalES03}, we have the following

\begin{lemma}[\cite{DBLP:conf/sigmod/AgrawalES03}]
    \label{lemma:sec-previous-work}
    For polynomial $n$, the following distributions of $2\times n$ tuple is computationally indistinguishable with random sampled $a$ for all $\hat{x}_i$, and $X=\mathsf{HR}(\hat{X})$:
    \begin{align*}
        \{\left( x_1^a, ..., x_n^a\right)\}_{X, \lambda}\cindist \{\left(r_1, ..., r_n\right)\}_{X, \lambda}
    \end{align*}
    where each $r_i$ is independently and randomly sampled.
\end{lemma}
Next, we prove Lemma~\ref{lemma:security}.
\begin{proof}
We construct $\simulator_{\{\mathsf{send}, \mathsf{recv}\}}$ as follows:

\renewcommand{\descriptionlabel}[1]{\hspace{\labelsep}#1}
\begin{description}[style=nextline]
    \item[Simulator for sender:]
    Let $\simulator_\mathsf{send}$ first samples a uniform random key $b_r\sample \ZZ_p$ and a random permutation $\pi_r\in S_n$. 
    Then the simulator extracts the {leakage} $|I^{ab}_\mathsf{sub}|
    \triangleq X^{ab}_{\pi}\cap {Y^{ab}_\mathsf{sub}}$ $\tilde{n}$ (this leakage equals to the intersection size between sender's input and Receiver's sub-sampled input).
    To simulate the view, observe that in Step~3 of the DPSI protocol in Fig \ref{fig:dp-psi}, sender receives two messages from Receiver, i.e., $X_\pi^{ab}$, and $Y_\mathsf{sub}^b$.
    Simulator $\simulator_\mathsf{send}$ simulates those two messages with $R_1\sample\GG^n, R_2\sample\GG^{\tilde{n}}$.
    To finish the simulation, we need to show that for all $X, Y$,
    \begin{align}
        \begin{split}
            \{(R_1, R_2)\}_{X, Y, \lambda} \cindist
            \{(X_\pi^{ab},Y_\mathsf{sub}^b)\}_{X, Y, \lambda}
        \end{split}
        \label{eq:indist-send}
    \end{align}
    where $X_\pi^{ab}=\pi(\mathsf{HR}(X))^{ab}$ and $Y_\mathsf{sub}^{b}=\mathsf{HR}(Y_\mathsf{sub})^{b}$, here we use $\mathsf{HR}$ to denote the function for ``hash and re-order lexicographically''.
\end{description}

Now, we want to prove the two probablistic ensembles in the above Equation (\ref{eq:indist-send})
are indistinguishable. 

% Since we allow the simulator to extract the ``approximate'' size of the subsampled set (a.k.a. $|B^b_\mathsf{sub}|$), this could be easily proved with the lemma from \cite{DBLP:conf/sigmod/AgrawalES03}

To see the indistinguishability of Eq.~\ref{eq:indist-send}, we first let $X_1 = \pi(X^a)\in\GG^n$, $X_2 = Y_\mathsf{sub}\in\GG^{\tilde{n}}$, then by lemma \ref{lemma:sec-previous-work}, we have the following,
\begin{align*}
    \{\left(\mathsf{HR}(X_1)^b\|\mathsf{HR}(X_2)^b\right)\}\cindist\{\left(r_1, ..., r_{n+\tilde{n}}\right)\}= \{R_1, R_2\}
\end{align*}
where $\|$ means concatenation.

\begin{description}[style=nextline]
    \item[Simulator for Receiver:] Since only receives a single message from sender besides the final result, first let $\simulator_\mathsf{recv}$ extract the size of sender's input $n$. Then it samples a uniform random set $R\sample\GG^n$.
    That is, we need to show the following indistinguishability,
    \begin{align}
        \{R\}_{X, Y, \lambda} \cindist\{\mathsf{HR}(X)^a\}_{X, Y, \lambda}
    \end{align}
\end{description}

\noindent
Since $a$ is uniformly sampled, this indistinguishably could be directly proved by Lemma \ref{lemma:sec-previous-work}.

\end{proof}

\subsection{Proof of Lemma~\ref{claim:eq_card1_card2}}
\label{app:1st-equ}
\begin{IEEEproof}
To facilitate the DP analysis,   we represent  $\operatorname{Ber}(\pB)
\land \hash(y_i)^{ab}\in A^{ab}_{\pi}$ in Algorithm~\ref{algo:real-partical-protocol} in an equivalent compound expression as follows: 
\begin{equation}
\label{eq:logic-eqv}
    \begin{aligned}
    \operatorname{Ber}(\pB)
\land \hash(y_i)^{ab}\in A^{ab}_{\pi}
=&
\operatorname{Ber}(\pB)
\land  y_i\in A
\\
  =&\left(\neg \operatorname{Ber}(2(1-\pB)) \land y_i\in A\right)\lor \left(\operatorname{Ber}(2(1-\pB))\land  \mathrm{Ber(1/2)} \land y_i\in A\right),
\end{aligned}
\end{equation}
{where the symbols $\neg$, $\land$ and $\lor$ denote   logical negation, conjunction, and disjunction. The first equality in (\ref{eq:logic-eqv}) is due to the mapping between ciphertext and plaintext,} and the second equality is due to the fact that
\begin{equation*}
\begin{aligned}
p = & (1-2(1-\pB))+2(1-\pB)\times \frac{1}{2}\\
 =&   \Pr[\mathrm{Ber}(2(1-\pB))=0]+\Pr[\mathrm{ Ber}(2(1-\pB))=1]\cdot \Pr[\mathrm{Ber}(1/2)=1].
    \end{aligned}
\end{equation*}
Let $\mathsf{G}$  denote  the collection of successful events for the first Bernoulli trial, $\mathrm{Ber}(2(1-\pB))$, in (\ref{eq:logic-eqv}), and $G$ be a constant of the random set $\mathsf G$.
Then the set $\{i|i\notin G\}$ is the collection of failed events  for  this Bernoulli trial. 
For the PMF     $\mathcal M_{\mathsf{B}}(A, B)$ output in Algorithm~\ref{algo:real-partical-protocol},  we have

\begin{equation}
    \begin{aligned}
\Pr\left[|\mathsf I^{ab}_{\mathsf{sub}}|=z\right]
% =& \sum_{G \subseteq[n]} \Pr\left[|\mathsf I^{ab}_{\mathsf{sub}}| =z\cap \mathsf{G}=G\right] \\
=& \sum_{G \subseteq[n]} \Pr\left[\sum_{i=1}^n \mathsf{y_i}=z \Big| \mathsf{G}=G\right]
\Pr\left[ \mathsf{G}=G\right]\\
% =& \sum_{G \subseteq[n]} \Pr\left[|G\cap A|=\Psi| \mathsf{G}=G\right]
% \Pr\left[ \mathsf{G}=G\right]\\
\stackrel{(a)}=& \sum_{G \subseteq[n]} \Pr\left[\sum_{\substack{i \notin G,\\ \hash(y_i)^{ab}\in A^{ab}_{\pi}}} 1+\sum_{\substack{i \in G,\\ \hash(y_i)^{ab}\in A^{ab}_{\pi}}} \operatorname{Ber}\left(\frac{1}{2}\right)=z\right] \\
&\qquad \cdot\left(2(1-\pB)\right)^{|G|}\left(1-2(1-\pB)\right)^{n-|G|}\\
% =& \sum_{G \subseteq[n]} \Pr\left[\Big|([n]\setminus G)\cap \{i:y_i\in I\}\Big|  +\sum_{\{i \in G\} \cap \{i:y_i\in I\}} \operatorname{Ber}\left(\frac{1}{2}\right)=z\right] \\
% & \cdot\left(\frac{\lambda}{n}\right)^{|G|}\left(1-\frac{\lambda}{n}\right)^{n-|G|}\\
\stackrel{(b)}=& \sum_{G \subseteq[n]} \Pr\left[\left|I\setminus G\right|  +\sum_{\{i \in G\} \cap \{i:y_i\in I\}} \operatorname{Ber}\left(\frac{1}{2}\right)=z\right] 
  \left(2(1-\pB)\right)^{|G|}\left(1-2(1-\pB)\right)^{n-|G|}.
\end{aligned}
\label{eq:p_imaginary}
\end{equation}
 Eq.~$(a)$ follows (\ref{eq:logic-eqv}) by equivalently representing Line~2 in Algorithm~\ref{algo:real-partical-protocol}, and Eq.~$(b)$ is {due to the fact that $\hash(y_i)^{ab}\in A^{ab}_{\pi}$ implies $y_i\in I$}. 
Next, we prove that the PMF of output in Algorithm~\ref{algo:observe_card2}  is the same with the  PMF of Algorithm~\ref{algo:real-partical-protocol} in Eq.~(\ref{eq:p_imaginary}).
Let the random  $\mathsf T$ take a value  $T\subseteq B$ in Algorithm~\ref{algo:observe_card2}.
According to the law of total probability, we have
\begin{equation}
    \begin{aligned}
\Pr\left[|\mathsf I^\prime_{\mathsf{sub}}|=z\right] 
% =&\sum_{T \subseteq B} \Pr\left[\mathsf{z}=z \cap \mathsf{T}=T\right] \\
=&\sum_{T \subseteq B} \Pr\left[|\mathsf I^\prime_{\mathsf{sub}}|=z |\mathsf{T}=T\right]
\Pr\left[\mathsf{T}=T\right]\\
% \stackrel{(*)}=&\sum_{T \subseteq B} \Pr\left[\Big|(B\setminus T)\cap I\Big|+\sum_{y_i\in T_1}\operatorname{Ber}\left(\frac{1}{2}\right)=\Psi\right]\\
% &\quad \cdot  {n \choose |T|} \left(\frac{\lambda}{n}\right)^{|T|}\left(1-\frac{\lambda}{n}\right)^{n-|T|} \frac{1}{{n \choose |T|}}\\
\stackrel{(*)}=&\sum_{T \subseteq B} \Pr\left[\left|I\setminus T\right|+\sum_{y_i\in T\cap I}\operatorname{Ber}\left({1}/{2}\right)=z\right]\\
&\quad \cdot   
\operatorname{C}_n^{|T|}
\left(2(1-\pB)\right)^{|T|}\left(1-2(1-\pB)\right)^{n-|T|} \left(\operatorname{C}_n^{|T|}\right)^{-1}\\
=&\sum_{T \subseteq B} \Pr\left[\left|I\setminus T\right|+\sum_{y_i\in T\cap I}\operatorname{Ber}\left({1}/{2}\right)=z\right]\\
&\quad \cdot   \left(2(1-\pB)\right)^{|T|}\left(1-2(1-\pB)\right)^{n-|T|},
\end{aligned}
\label{eq:p_prime}
\end{equation}
where Eq. $(*)$    follows the procedure in Algorithm~\ref{algo:observe_card2} by sampling $|T|$ via $\operatorname{Bin}\left(n,2(1-\pB)\right)$ in $B$ and then  sampling from $\mathcal{T}_{s}$ uniformly with a constant cardinality $|T|$.
We thus have shown that
Eq. (\ref{eq:p_prime}) and  (\ref{eq:p_imaginary}) are equal, because the former is the marginalization of all subsets of Bob's data, and equivalently the latter is just the marginalization of all subsets of all the indices of Bob's data.
\end{IEEEproof}

\subsection{Proof of Lemma~\ref{claim:eq_card2_card3} }
\label{app:2nd-equ}

\begin{IEEEproof} 
According to the total probability, for $|\mathsf I^{\prime\prime}_{\mathsf{sub}}|$ in Algorithm~\ref{algo:observe_card3}, we have
\begin{equation}
\begin{aligned}
    \Pr\left[|\mathsf I^{\prime\prime}_{\mathsf{sub}}|=z\right] = \sum_{T_1 \subseteq I}& \Pr\left[\Big|I\setminus T_1\Big|+\sum_{y_i\in T_1}\operatorname{Ber}\left({1}/{2}\right)=z\right]\cdot   \left(2(1-\pB)\right)^{|T_1|}\left(1-2(1-\pB)\right)^{|I|-|T_1|}.
\end{aligned}
\label{eq:z1}
\end{equation}
On the other hand, according to (\ref{eq:p_prime}), for $\mathcal M^{\prime}_{\mathsf{int}}(A, B)$ in Algorithm~\ref{algo:observe_card2}, we have
\begin{equation*}
    \begin{aligned}
\Pr\left[|\mathsf I^\prime_{\mathsf{sub}}|=z{\color{blue}}\right] 
=&\sum_{T \subseteq B} \Pr\left[\left|I\setminus T\right|+\sum_{y_i\in T_1}\operatorname{Ber}\left({1}/{2}\right)=z\right]
\cdot   \left(2(1-\pB)\right)^{|T|}\left(1-2(1-\pB)\right)^{n-|T|},\\
% =&\sum_{T_2  \subseteq B\setminus I}\sum_{T_1  \subseteq I} \Pr\left[\left|I\setminus T\right|+\sum_{y_i\in T_1}\operatorname{Ber}\left(\frac{1}{2}\right)=z\right]\\
% &\quad \cdot   \left(2(1-\pB)\right)^{|T_1|+|T_2|}\left(1-2(1-\pB)\right)^{n-|T_1|-|T_2|},\\
\stackrel{(a)}=&\sum_{T_2  \subseteq B\setminus I}\sum_{T_1  \subseteq I} \Pr\left[\left|I\setminus T_1\right|+\sum_{y_i\in T_1}\operatorname{Ber}\left({1}/{2}\right)=z\right]   \left(2(1-\pB)\right)^{|T_1|+|T_2|}\left(1-2(1-\pB)\right)^{n-|T_1|-|T_2|}\\
\stackrel{(b)}=&\sum_{T_2  \subseteq B\setminus I} \left(2(1-\pB)\right)^{|T_2|} \left(1-2(1-\pB)\right)^{n-|I|-|T_2|} \Bigg(
\sum_{T_1  \subseteq I} \Pr\left[\left|I\setminus T_1\right|+\sum_{y_i\in T_1}\operatorname{Ber}\left({1}/{2}\right)=z\right]\\
&\quad\cdot   \left(2(1-\pB)\right)^{|T_1|}\left(1-2(1-\pB)\right)^{I-|T_1|}\Bigg),\\
\stackrel{(c)}=& 
\Pr[|\mathsf I^{\prime\prime}_{\mathsf{sub}}| = z]\sum_{T_2  \subseteq B\setminus I} \left(2(1-\pB)\right)^{|T_2|} \left(1-2(1-\pB)\right)^{n-|I|-|T_2|},%\\
% =&\kappa \sum_{T_1 \subseteq I} \Pr\left[\Big|I\setminus T_1\Big|+\sum_{y_i\in T_1}\operatorname{Ber}\left(\frac{1}{2}\right)=z\right]\\
% &\quad \cdot   \left(2(1-\pB)\right)^{|T_1|}\left(1-2(1-\pB)\right)^{|I|-|T_1|},
\end{aligned}
\end{equation*}
where $(a)$ follows from letting $T_2 = T\setminus T_1$,  $(b)$ can be obtained by factoring out the terms that does not depend on $T_1$, and $(c)$ is achieved by plugging in (\ref{eq:z1}) as well as the proof of the  in the following: 
\begin{equation*}
    \begin{aligned}
   & \sum_{T_2  \subseteq B\setminus I} \left(2(1-\pB)\right)^{|T_2|} \left(1-2(1-\pB)\right)^{n-|I|-|T_2|}\\
 \stackrel{(a)}   = & \left(1-2(1-\pB)\right)^{n-|I|} \cdot \sum_{T_2  \subseteq B\setminus I} \left(\frac{2(1-\pB)}{1-2(1-\pB)}\right)^{|T_2|}\\
=& \left(1-2(1-\pB)\right)^{n-|I|} \cdot \sum_{c=0}^{|B\setminus I|} {|B\setminus I| \choose c}  \left(\frac{2(1-\pB)}{1-2(1-\pB)}\right)^c\\
=& \left(1-2(1-\pB)\right)^{n-|I|} \cdot \sum_{c=0}^{|B\setminus I|} {|B\setminus I| \choose c}  \left(\frac{2(1-\pB)}{1-2(1-\pB)}\right)^c 1^{|B\setminus I|-c}\\
 \stackrel{(b)}=& \left(1-2(1-\pB)\right)^{n-|I|} \cdot \left(1+\frac{2(1-\pB)}{1-2(1-\pB)}\right)^{|B\setminus I|}\\
 \stackrel{(c)} =&\left(1-2(1-\pB)\right)^{n-|I|} \cdot \left(\frac{1}{1-2(1-\pB)}\right)^{n-|I|}\\
 =&1,
    \end{aligned}
\end{equation*}
where $(a)$ follows by factoring out terms that are independent of $T_2$,   $(b)$ is due to the binomial expansion, $(c)$ is because $n-|I| = |B\setminus I|$.  Hence, we complete the proof. 
\end{IEEEproof}

\subsection{Proof of Lemma~\ref{claim:C_T_dp}}
\label{app:1st-random}
\begin{IEEEproof} First, %since %$\operatorname{Bin}(|H|,\frac{1}{2}) = \sum_{h\in H}b_h$, where
% $b_j\sim\operatorname{Ber}(\frac{1}{2})$,   $b_j\in[0,1]$, and   $\mathbb{E}[b_j] = \frac{1}{2}$. Then, 
by applying the Hoeffding's inequality to the $\operatorname{Ber}(\frac{1}{2})$ for all $y_i\in T_1$ in Line~1 of Algorithm~\ref{algo:decomp1},  we obtain
% \begin{equation*}
%     \Pr\left(\left|\sum_{h\in J}b_j-\frac{|J|}{2}\right|<\sqrt{\frac{|J|}{2}\log\frac{4}{\delta}}\right) > 1-\frac{\delta}{2}.
% \end{equation*}
\begin{equation*}
    \Pr\left[\left|\sum_{y_i\in T_1}\operatorname{Ber}\left(\frac{1}{2}\right)-\frac{|T_1|}{2}\right|<\sqrt{\frac{|T_1|}{2}\log\frac{4}{\delta}}\right] > 1-\frac{\delta}{2}.
\end{equation*}
We reformulate the above inequality by
$$\Pr\left[\sum_{y_i\in T_1}\operatorname{Ber}\left(\frac{1}{2}\right)\notin I_u\right]<\frac{\delta}{2}$$
with
\begin{equation}\label{eq:ber_sum_range}
    I_u\triangleq \left(\frac{|T_1|}{2}-\sqrt{\frac{|T_1|}{2}\log\frac{4}{\delta}},\frac{|T_1|}{2}+\sqrt{\frac{|T_1|}{2}\log\frac{4}{\delta}}\right).
\end{equation}
Furthermore, following the DP definition, for any $ W\subseteq \{0,1,\ldots,|I|\}$, we have
\begin{equation*}
\begin{aligned}
    &\Pr[\mathsf{z}_1(B)\in W] \\
    =& \Pr\left[\left\{\mathsf{z}_1(B)\in W\right\} \cap \left\{\sum_{y_i\in T_1}\operatorname{Ber}\left(\frac{1}{2}\right)\in I_u\right\}\right] 
    + \Pr\left[\left\{\mathsf{z}_1(B)\in W\right\} \cap \left\{\sum_{y_i\in T_1}\operatorname{Ber}\left(\frac{1}{2}\right)\notin I_u\right\}\right]\\
  \leq&   \Pr\left[\left\{\mathsf{z}_1(B)\in W\right\} \cap \left\{\sum_{y_i\in T_1}\operatorname{Ber}\left(\frac{1}{2}\right)\in I_u\right\}\right]  + \frac{\delta}{2}\\
%   = &\sum_{r\in W\cap I_u} \Pr\left(B+\sum^n_{i\notin H,\hash(y_i)^{ab}\in A^{ab}_{\pi}}1=r\right)+  \frac{\delta}{2}\\
  = &\sum_{w\in W} \Pr\left[\left(\Big|I\setminus T_1\Big|+\sum_{y_i\in T_1}\operatorname{Ber}\left(\frac{1}{2}\right)\right)=w\right]+ \frac{\delta}{2},
\end{aligned}
\end{equation*}
with $\sum_{y_i\in T_1}\operatorname{Ber}\left(\frac{1}{2}\right)\in I_u$.

% \begin{equation*}
%     \begin{aligned}
% \frac{\Pr_{\mathsf{z}_1}(\mathsf{z}_1(B)\in W)-\frac{\delta}{2}}{\Pr_{\mathsf{z}_1}(\mathsf{z}_1(\widetilde{B})\in W)}\leq \exp(\epsilon)
%     \end{aligned}
% \end{equation*}
Let $\widetilde{B}$ be any neighboring dataset of $B$ that they differs by one data record. Moreover, define $\widetilde I \triangleq A\cap \widetilde B$ and $\widetilde T_1\subset \widetilde I $ be the counterpart of $T$ and $T_1$ in Algorithm~\ref{algo:decomp1}, respectively. Thus, to complete the proof of $(\epsilon,\frac{1}{2}\delta)$-DP, we need to further to prove that for any $T_1$, $\widetilde T_1$, $I$ and $\widetilde I$, the following holds % and \hl{replace$r\in W\cap I_u$},

% $\frac{ \Pr[C_H(X)\in W]}{ \Pr[C_H(X')\in W]}\leq \frac{upper bound2}{upper bound2} \leq1+\epsilon +\delta$

% % $ \Pr[C_H(X)\in W] \leq $

% \hl{ $\Pr[C_H(X)\in W] \leq  (1+\epsilon)\Pr[C_H(X')\in W] +\delta$}

% we have $\Pr[C_H(X)\in W] \leq\sum \Pr +\frac{\delta}{2}$

% require $\sum \Pr \leq  (1+\epsilon)\Pr[C_H(X')\in W]$

% known $ (1+\epsilon)\Pr[C_H(X')\in W] \leq \sum (1+\epsilon) \Pr'$

\begin{equation}\label{eq:general_requirement0}
\frac{\sum_{w\in W}\Pr\left[\left(\Big|I\setminus T_1\Big|+\sum_{y_i\in T_1}\operatorname{Ber}\left(\frac{1}{2}\right)\right)=w\right]}{\sum_{w\in W} \Pr\left[\left(\Big|\widetilde{I}\setminus \widetilde{T}_1\Big|+\sum_{y_i\in \widetilde{T}_1}\operatorname{Ber}\left(\frac{1}{2}\right)\right)=w\right]} \leq e^{\epsilon_{\mathsf{z}_1}}.
\end{equation}
Due to the facts that 
$1+\eA \leq \exp\left(x\right)$ for any $x\geq 0$ and
\begin{equation}
\begin{split}
&\frac{\sum_{w\in W}\Pr\left[\left(\Big|I\setminus T_1\Big|+\sum_{y_i\in T_1}\operatorname{Ber}\left(\frac{1}{2}\right)\right)=w\right]}{\sum_{w\in W} \Pr\left[\left(\Big|\widetilde{I}\setminus \widetilde{T}_1\Big|+\sum_{y_i\in \widetilde{T}_1}\operatorname{Ber}\left(\frac{1}{2}\right)\right)=w\right]}\\ \leq &
\max_{w\in W}\left\{\frac{\Pr\left[\sum_{y_i\in T_1}\operatorname{Ber}\left(\frac{1}{2}\right)=w-\Big|I\setminus T_1\Big|\right]}{\Pr\left[\sum_{y_i\in \widetilde{T}_1}\operatorname{Ber}\left(\frac{1}{2}\right)=w-\Big|\widetilde{I}\setminus \widetilde{T}_1\Big|\right]}\right\},
\end{split}
\end{equation}
it is sufficient to prove the holding of the following inequality for the $(\epsilon_{\mathsf{z}_1},\frac{1}{2}\delta)$-DP guarantee:
\begin{equation}\label{eq:general_requirement}
\max\left\{\frac{\Pr\left[\sum_{y_i\in T_1}\operatorname{Ber}\left(\frac{1}{2}\right)=w-\Big|I\setminus T_1\Big|\right]}{\Pr\left[\sum_{y_i\in \widetilde{T}_1}\operatorname{Ber}\left(\frac{1}{2}\right)=w-\Big|\widetilde{I}\setminus \widetilde{T}_1\Big|\right]}\right\}
\leq 1+\epsilon_{\mathsf{z}_1},
\end{equation}
with $\sum_{y_i\in T_1}\operatorname{Ber}\left(\frac{1}{2}\right)\in I_u$ and $\sum_{y_i\in \widetilde T_1}\operatorname{Ber}\left(\frac{1}{2}\right)\in I_u$.

Without loss of generality, assume that $B$ and $\widetilde B$ differs by the $j$th data record, i.e., $b_j\neq \widetilde b_j$ with $b_j\in B$ and $\widetilde b_j\in B$.  We  partition $B$ ($\widetilde B$) into three non-overlapping subsets, i.e., $T_1$ ($\widetilde T_1$), $I\setminus T_1$ ($\widetilde I\setminus \widetilde T_1$), and $\widetilde B\setminus I$ ($\widetilde B\setminus \widetilde I$). 
Since the only different pair of element $b_j$ and $\widetilde b_j$ could belong to any of these three subsets, respectively, there are in total nine different cases to analyze for the l.h.s. of  (\ref{eq:general_requirement}) as following.
% \begin{figure}[htbp]
% \centering
% \includegraphics[width=1\linewidth]{images/B_cases.png}
% \caption{....}
% \label{fig:B-cases}
% \end{figure}

\noindent\textbf{Case 1.}  If $b_j\in  B\setminus I$ and $\widetilde b_j\in \widetilde I\setminus \widetilde T_1$, following the neighboring data sets definition,  we have $|I\setminus T_1| +1 = |\widetilde  I\setminus \widetilde  T_1|$. Substitute the result to the l.h.s of (\ref{eq:general_requirement}), we obtain 
\begin{equation}\label{eq:C_T_1_proof}
\begin{aligned}
% & \frac{\Pr\left[\left(\Big|I\setminus T_1\Big|+\sum_{y_i\in T_1}\operatorname{Ber}\left(\frac{1}{2}\right)\right)=w\right]}{\Pr\left[\left(\Big|\widetilde{I}\setminus \widetilde{T}_1\Big|+\sum_{y_i\in T_1}\operatorname{Ber}\left(\frac{1}{2}\right)\right)=w\right]} \\
\mathrm{l.h.s.\ of\ (\ref{eq:general_requirement})} 
% = & \frac{\Pr\left[\sum_{y_i\in T_1}\operatorname{Ber}\left(\frac{1}{2}\right)=w-\Big|I\setminus T_1\Big|\right]}{\Pr\left[\sum_{y_i\in \widetilde{T}_1}\operatorname{Ber}\left(\frac{1}{2}\right)=w-\Big|\widetilde{I}\setminus \widetilde{T}_1\Big|\right]}\\
= & \frac{\Pr\left[\sum_{y_i\in T_1}\operatorname{Ber}\left(\frac{1}{2}\right)=w-\Big|I \setminus  {T}_1\Big|\right]}{\Pr\left[\sum_{y_i\in \widetilde{T}_1}\operatorname{Ber}\left(\frac{1}{2}\right)=w-\Big|{I}\setminus {T}_1\Big|-1\right]}\\
%   =& \frac{\Pr\left(\left[\sum_{y_i\in T_1}\operatorname{Ber}\left(\frac{1}{2}\right)\right]=w-\sum_{y_i\in(\widetilde{B}/T)\cap I}1+1\right)}{\Pr\left(\left[\sum_{y_i\in T_1}\operatorname{Ber}\left(\frac{1}{2}\right)\right]=w-\sum_{y_i\in(\widetilde{B}/T)\cap I}1\right)}\\
 = &\frac{\Pr\left[\sum_{y_i\in T_1}\operatorname{Ber}\left(\frac{1}{2}\right)=k\right]}{\Pr\left[\sum_{y_i\in \widetilde{T}_1}\operatorname{Ber}\left(\frac{1}{2}\right)=k-1\right]} \qquad (\mathrm{let\ }k = w-|I\setminus T_1|)\\
 \stackrel{(*)}    = & \frac{ {|T_1| \choose k} (1/2)^{k} (1/2)^{|T_1|-k} }{{|T_1| \choose k-1} (1/2)^{k-1} (1/2)^{|T_1|-k+1}}\\
    =& \frac{|T_1|-k+1}{k}\\
     < & \frac{\frac{|T_1|}{2}+\sqrt{\frac{|T_1|}{2}\log\frac{4}{\delta}}+1}{\frac{|T_1|}{2}-\sqrt{\frac{|T_1|}{2}\log\frac{4}{\delta}}},
    \end{aligned}
\end{equation}
where  $(*)$ is due to the fact that 
the summation of mutually independent Bernoulli random variables can be seen as a binomial distribution 
and $|T_1| = |\widetilde{T}_1|$ in this case.
The last inequality is because that $k=\sum_{y_i\in T_1}\operatorname{Ber}\left(\frac{1}{2}\right)$ belongs to $I_u$ with lower and upper bound defined in Eq.~(\ref{eq:ber_sum_range}). We take the value $k\in I_u$ in this range to obtain the maximum value  for (\ref{eq:general_requirement}).
For short, we define
\begin{equation}
\label{eq:eps1}
1+\epsilon_{1,\mathsf{z}_1}
\triangleq 
\frac{\frac{|T_1|}{2}+\sqrt{\frac{|T_1|}{2}\log\frac{4}{\delta}}+1}{\frac{|T_1|}{2}-\sqrt{\frac{|T_1|}{2}\log\frac{4}{\delta}}}.
\end{equation}

% , $(c)$ is because $k\in I_u$ (see (\ref{eq:ber_sum_range})), $(d)$ is achieved by defining $u = \sqrt{\frac{|T_1|}{2}\log\frac{4}{\delta}}$, and finally $(e)$ holds as long as $|T_1|\geq 18\log \frac{4}{\delta}$. Finally, we can complete the proof by setting $\epsilon = \sqrt{\frac{18\log\frac{4}{\delta}}{|T_1|}}.$

\noindent\textbf{Case 2.} If $b_j\in  B\setminus I$ and $\widetilde b_j\in \widetilde B\setminus \widetilde I$,   
following the neighboring data sets definition,  we have $|I\setminus T_1| = |\widetilde  I\setminus \widetilde  T_1|$ and $|T_1 |=\widetilde  T_1|$. Since all elements in $T_1$ and $\widetilde  T_1$ are subject to independent Bernoulli trials, the l.h.s. of (\ref{eq:general_requirement}) is $1$, which makes the $(\epsilon_{\mathsf{z}_1}, \frac{1}{2}\delta)$-DP always hold for any $\epsilon_{\mathsf{z}_1}>0$ in this case. 

\noindent\textbf{Case 3.} If $b_j\in  B\setminus I$ and $\widetilde b_j\in \widetilde T_1$,   
following the neighboring data sets definition, we reach   $|I\setminus T_1| = |\widetilde  I\setminus \widetilde  T_1|$ and $|T_1|+1=|\widetilde  T_1|$. Thus,
\begin{equation*}\label{eq:C_T_1_proof}
\begin{aligned}
\mathrm{l.h.s.\ of\ (\ref{eq:general_requirement})} = & \frac{\Pr\left[\sum_{y_i\in T_1}\operatorname{Ber}\left(\frac{1}{2}\right)=w-\Big|I\setminus T_1\Big|\right]}{\Pr\left[\sum_{y_i\in \widetilde{T}_1}\operatorname{Ber}\left(\frac{1}{2}\right)=w-\Big|{I}\setminus {T}_1\Big|\right]}\\
 = &\frac{\Pr\left[\sum_{y_i\in T_1}\operatorname{Ber}\left(\frac{1}{2}\right)=k\right]}{\Pr\left[\sum_{y_i\in \widetilde{T}_1}\operatorname{Ber}\left(\frac{1}{2}\right)=k\right]} \qquad (\mathrm{let\ }k = w-|I\setminus T_1|)\\
    = & \frac{ {|T_1| \choose k}  }{{|T_1| +1 \choose k} } = \frac{|T_1|+1-k}{|T_1|+1}< 1,
    \end{aligned}
\end{equation*}
which also makes (\ref{eq:general_requirement}) always hold in this case.

\noindent\textbf{Case 4.} If $b_j\in I\setminus T_1$ and $\widetilde b_j\in \widetilde I\setminus \widetilde T_1$, then we have $|I\setminus T_1| = |\widetilde  I\setminus \widetilde  T_1|$ and $|T_1|$ = $|\widetilde  T_1|$, which is similar to Case 2 that  the $(\epsilon, \frac{1}{2}\delta)$-DP always hold in this case.

\noindent\textbf{Case 5.} If $b_j\in I\setminus T_1$ and $\widetilde b_j\in \widetilde B\setminus \widetilde I$, following the neighboring data sets definition,  $|I\setminus T_1| -1 = |\widetilde  I\setminus \widetilde  T_1|$ and $|T_1|$ = $|\widetilde  T_1|$. Thus, 
\begin{equation*}\label{eq:C_T_1_proof}
\begin{aligned}
\mathrm{l.h.s.\ of\ (\ref{eq:general_requirement})} 
% = & \frac{\Pr\left[\sum_{y_i\in T_1}\operatorname{Ber}\left(\frac{1}{2}\right)=w-\Big|I\setminus T_1\Big|\right]}{\Pr\left[\sum_{y_i\in \widetilde{T}_1}\operatorname{Ber}\left(\frac{1}{2}\right)=w-\Big|\widetilde{I}\setminus \widetilde{T}_1\Big|\right]}\\
= & \frac{\Pr\left[\sum_{y_i\in T_1}\operatorname{Ber}\left(\frac{1}{2}\right)=w-\Big|{I}\setminus {T}_1\Big|\right]}{\Pr\left[\sum_{y_i\in \widetilde{T}_1}\operatorname{Ber}\left(\frac{1}{2}\right)=w-\Big|{I}\setminus {T}_1\Big|+1\right]}\\
 = &\frac{\Pr\left[\sum_{y_i\in T_1}\operatorname{Ber}\left(\frac{1}{2}\right)=k\right]}{\Pr\left[\sum_{y_i\in \widetilde{T}_1}\operatorname{Ber}\left(\frac{1}{2}\right)=k+1\right]} \qquad (\mathrm{let\ }k = w-|I\setminus T_1|)\\
  = & \frac{ {|T_1| \choose k}  }{{|T_1| \choose k+1} }= \frac{k+1}{|T_1|-k}
         <  \frac{\frac{|T_1|}{2}+\sqrt{\frac{|T_1|}{2}\log\frac{4}{\delta}}+1}{\frac{|T_1|}{2}-\sqrt{\frac{|T_1|}{2}\log\frac{4}{\delta}}} \quad(k\in I_u).
%   \stackrel{(c)}   \leq & \frac{\frac{|T_1|}{2}+\sqrt{\frac{|T_1|}{2}\log\frac{4}{\delta}}}{\frac{|T_1|}{2}-\sqrt{\frac{|T_1|}{2}\log\frac{4}{\delta}}+1} \\ %\qquad (k\in I_u)\\
%     < & \frac{\frac{|T_1|}{2}+\sqrt{\frac{|T_1|}{2}\log\frac{4}{\delta}}}{\frac{|T_1|}{2}-\sqrt{\frac{|T_1|}{2}\log\frac{4}{\delta}}},
    \end{aligned}
\end{equation*}
We denote 
\begin{equation}
\label{eq:eps5}
1+\epsilon_{5,\mathsf{z}_1} \triangleq \frac{\frac{|T_1|}{2}+\sqrt{\frac{|T_1|}{2}\log\frac{4}{\delta}}+1}{\frac{|T_1|}{2}-\sqrt{\frac{|T_1|}{2}\log\frac{4}{\delta}}}.
\end{equation}
Then 
\begin{equation}
\label{eq:eps5-2}
\epsilon_{5,\mathsf{z}_1} = \frac{2\sqrt{\frac{|T_1|}{2}\log\frac{4}{\delta}}+1}{\frac{|T_1|}{2}-\sqrt{\frac{|T_1|}{2}\log\frac{4}{\delta}}}.
\end{equation}
\noindent\textbf{Case 6.} If $b_j\in I\setminus T_1$ and $\widetilde b_j \in \widetilde T_1$, following the neighboring data sets definition,  $|I\setminus T_1|-1 = |\widetilde  I\setminus \widetilde  T_1| $ and $|T_1| +1= |\widetilde  T_1| $. Thus, we obtain the followings
\begin{equation*}\label{eq:C_T_1_proof}
\begin{aligned}
\mathrm{l.h.s.\ of\ (\ref{eq:general_requirement})} 
% = & \frac{\Pr\left[\sum_{y_i\in T_1}\operatorname{Ber}\left(\frac{1}{2}\right)=w-\Big|I\setminus T_1\Big|\right]}{\Pr\left[\sum_{y_i\in \widetilde{T}_1}\operatorname{Ber}\left(\frac{1}{2}\right)=w-\Big|\widetilde{I}\setminus \widetilde{T}_1\Big|\right]}\\
= & \frac{\Pr\left[\sum_{y_i\in T_1}\operatorname{Ber}\left(\frac{1}{2}\right)=w-\Big|{I}\setminus {T}_1\Big|\right]}{\Pr\left[\sum_{y_i\in \widetilde{T}_1}\operatorname{Ber}\left(\frac{1}{2}\right)=w-\Big|{I}\setminus {T}_1\Big|+1\right]}\\
 \stackrel{(a)} = &\frac{\Pr\left[\sum_{y_i\in T_1}\operatorname{Ber}\left(\frac{1}{2}\right)=k\right]}{\Pr\left[\sum_{y_i\in \widetilde{T}_1}\operatorname{Ber}\left(\frac{1}{2}\right)=k+1\right]} \qquad (\mathrm{let\ }k = w-|I\setminus T_1|)\\
    = & \frac{ {|T_1| \choose k}   }{{|T_1|+1 \choose k +1} } = \frac{k+1}{|T_1|+1}<1,
    \end{aligned}
\end{equation*}
which   makes (\ref{eq:general_requirement}) trivially hold. 

\noindent\textbf{Case 7.} If $b_j\in T_1$ and $\widetilde b_j\in \widetilde I\setminus \widetilde T_1$, following the neighboring data sets definition $|I\setminus T_1| +1= |\widetilde I\setminus \widetilde T_1|$ and $|T_1|-1 = |\widetilde T_1|$. Thus, 
\begin{equation*}\label{eq:C_T_1_proof}
\begin{aligned}
\mathrm{l.h.s.\ of\ (\ref{eq:general_requirement})} 
% = & \frac{\Pr\left[\sum_{y_i\in T_1}\operatorname{Ber}\left(\frac{1}{2}\right)=w-\Big|I\setminus T_1\Big|\right]}{\Pr\left[\sum_{y_i\in \widetilde{T}_1}\operatorname{Ber}\left(\frac{1}{2}\right)=w-\Big|\widetilde{I}\setminus \widetilde{T}_1\Big|\right]}\\
= & \frac{\Pr\left[\sum_{y_i\in T_1}\operatorname{Ber}\left(\frac{1}{2}\right)=w-\Big| {I}\setminus {T}_1\Big|\right]}{\Pr\left[\sum_{y_i\in \widetilde{T}_1}\operatorname{Ber}\left(\frac{1}{2}\right)=w-\Big|{I}\setminus {T}_1\Big|-1\right]}\\
= &\frac{\Pr\left[\sum_{y_i\in T_1}\operatorname{Ber}\left(\frac{1}{2}\right)=k\right]}{\Pr\left[\sum_{y_i\in \widetilde{T}_1}\operatorname{Ber}\left(\frac{1}{2}\right)=k-1\right]} \qquad (\mathrm{let\ }k = w-\Big|{I}\setminus{T}_1\Big|)\\
    = & \frac{ {|T_1| \choose k}   }{{|T_1|-1 \choose k-1} }=\frac{|T_1|}{k}<1+\frac{\frac{|T_1|}{2}+\sqrt{\frac{|T_1|}{2}\log\frac{4}{\delta}}}{\frac{|T_1|}{2}-\sqrt{\frac{|T_1|}{2}\log\frac{4}{\delta}}} \quad(k\in I_u).
    \end{aligned}
\end{equation*}
For short, we define
\begin{equation}
\label{eq:eps7}
    1+\epsilon_{7,\mathsf{z}_1} \triangleq 1+\frac{\frac{|T_1|}{2}+\sqrt{\frac{|T_1|}{2}\log\frac{4}{\delta}}}{\frac{|T_1|}{2}-\sqrt{\frac{|T_1|}{2}\log\frac{4}{\delta}}}.
\end{equation}

\noindent\textbf{Case 8.} If $b_j\in T_1$ and $\widetilde b_j\in \widetilde B\setminus \widetilde I$, following the neighboring data sets definition $|I\setminus T_1| = |\widetilde I\setminus \widetilde T_1|$ and $|T_1| -1= |\widetilde T_1|$. Thus,
\begin{equation*}\label{eq:C_T_1_proof}
\begin{aligned}
\mathrm{l.h.s.\ of\ (\ref{eq:general_requirement})} 
% = & \frac{\Pr\left[\sum_{y_i\in T_1}\operatorname{Ber}\left(\frac{1}{2}\right)=w-\Big|I\setminus T_1\Big|\right]}{\Pr\left[\sum_{y_i\in \widetilde{T}_1}\operatorname{Ber}\left(\frac{1}{2}\right)=w-\Big|\widetilde{I}\setminus \widetilde{T}_1\Big|\right]}\\
= & \frac{\Pr\left[\sum_{y_i\in T_1}\operatorname{Ber}\left(\frac{1}{2}\right)=w-\Big|{I}\setminus {T}_1\Big|\right]}{\Pr\left[\sum_{y_i\in \widetilde{T}_1}\operatorname{Ber}\left(\frac{1}{2}\right)=w-\Big|{I}\setminus {T}_1\Big|\right]}\\
%   =& \frac{\Pr\left(\left[\sum_{y_i\in T_1}\operatorname{Ber}\left(\frac{1}{2}\right)\right]=w-\sum_{y_i\in(\widetilde{B}/T)\cap I}1+1\right)}{\Pr\left(\left[\sum_{y_i\in T_1}\operatorname{Ber}\left(\frac{1}{2}\right)\right]=w-\sum_{y_i\in(\widetilde{B}/T)\cap I}1\right)}\\
 = &\frac{\Pr\left[\sum_{y_i\in T_1}\operatorname{Ber}\left(\frac{1}{2}\right)=k\right]}{\Pr\left[\sum_{y_i\in \widetilde{T}_1}\operatorname{Ber}\left(\frac{1}{2}\right)=k\right]} \qquad (\mathrm{let\ }k = w-\Big|I\setminus {T}_1\Big|)\\
    = & \frac{ {|T_1| \choose k}   }{{|T_1|-1 \choose k} }=\frac{|T_1|}{|T_1|-k}\\
    \leq & 1+\frac{\frac{|T_1|}{2}-\sqrt{\frac{|T_1|}{2}\log\frac{4}{\delta}}}{\frac{|T_1|}{2}+\sqrt{\frac{|T_1|}{2}\log\frac{4}{\delta}}} \qquad (k\in I_u).
    \end{aligned}
\end{equation*}
For short, we denote
\begin{equation}
\label{eq:eps8}
 1+\epsilon_{8,\mathsf{z}_1} \triangleq 1+\frac{\frac{|T_1|}{2}-\sqrt{\frac{|T_1|}{2}\log\frac{4}{\delta}}}{\frac{|T_1|}{2}+\sqrt{\frac{|T_1|}{2}\log\frac{4}{\delta}}}.   
\end{equation}

\noindent\textbf{Case 9.} If $b_j\in T_1$ and $\widetilde b_j\in \widetilde T_1$,    $|I\setminus T_1| = |\widetilde  I\setminus \widetilde  T_1|$ and $|T_1|$ = $|\widetilde  T_1|$, which is similar to Case 2 and the $(\epsilon, \frac{1}{2}\delta)$-DP always hold for any $\epsilon>0$ in this case.

In summary, in the above Cases 2, 3, 4, 6 and 9,  $(\epsilon, \frac{1}{2}\delta)$-DP always hold for any $\epsilon, \delta>0$. Therefore, to study the worst case for DP guarantee, we focus on Cases 1, 5, 7 and 8 such that
\begin{equation*}
    \max\{1+\epsilon_1, 1+\epsilon_{5,\mathsf{z}_1}, 1+\epsilon_{7,\mathsf{z}_1}, 1+\epsilon_{8,\mathsf{z}_1}\} \leq 1+\epsilon_{\mathsf{z}_1}
\end{equation*}
to make sure Eq.~(\ref{eq:general_requirement}) always hold.
To obtain the express for $\epsilon_{\mathsf{z}_1}$, we need to compare $\epsilon_{1,\mathsf{z}_1}$, $\epsilon_{5,\mathsf{z}_1}$, $\epsilon_{7,\mathsf{z}_1}$, and $\epsilon_{8,\mathsf{z}_1}$ defined in (\ref{eq:eps1}), (\ref{eq:eps5}), (\ref{eq:eps7}), and (\ref{eq:eps8}), respectively.

{(1) When $\frac{|T_1|}{2}-\sqrt{\frac{|T_1|}{2}\log\frac{4}{\delta}}\leq 0$, it is evident that $\epsilon_{1,\mathsf{z}_1}$, $\epsilon_{5,\mathsf{z}_1}$, $\epsilon_{7,\mathsf{z}_1}$, and $\epsilon_{8,\mathsf{z}_1}$ are all smaller than $0$, and $(0,\delta)$-DP is guaranteed in this  scenario.}

(2) When $1>\frac{|T_1|}{2}-\sqrt{\frac{|T_1|}{2}\log\frac{4}{\delta}}>0$, which is equiavalent to 
%$ <|T_1|<2\log\frac{4}{\delta}$
{\begin{equation}
\label{eq:negligibleT1}
   2\log\frac{4}{\delta}< |T_1|< 2+\log\frac{4}{\delta}+\sqrt{4\log\frac{4}{\delta}+\left(\log\frac{4}{\delta}\right)^2}.
\end{equation}
In the following, we show that this scenario happens with negligible probability.    
%are always guaranteed in this case.
For simplicity, we denote the probability of $|T_1|$ lies in the range shown in (\ref{eq:negligibleT1}) as $\delta^\prime$, then we have
\begin{equation}
\label{eq:delta-prm}
\begin{split}
\delta^\prime =& \operatorname{F}\left(2+\log\frac{4}{\delta}+\sqrt{4\log\frac{4}{\delta}+\left(\log\frac{4}{\delta}\right)^2}\Big||I|,2(1-\pB)\right)\\ 
&- \operatorname{F}\left(2\log\frac{4}{\delta} \Big||I|,2(1-\pB)\right),
\end{split}
\end{equation}
where $\operatorname{F}(x|n, \pB)$ denotes the  Cumulative Distribution Function (CDF) of the Binomial distribution\footnote{CDF $\operatorname{F}(x|n,\pB)$ computes the cumulative probability of at most $x$ successful trials out of $n$ trials when the successful probability is  $p$.}. 

We show that    the first term  in Eq~(\ref{eq:delta-prm}), which is an upper bound of $\delta^{\prime}$
 is  negligible. Since the CDF of the Binomial distribution $\operatorname{F}(x|n,\pB)$ decreases with the total number of trials $n$ when $x$ and $p$ are fixed, we  consider the worse case of the first term by plugging  $|I|$ as small as possible. 
 In practice, we consider the relative large-scale problem with  $|I|\gg 200$. 
 For example, fix a sufficient small $\delta$, $\delta = \frac{1}{10^{10}}$ and set $p = \frac{e}{1+e}$,  if $|I| = 200$, the first term takes value of $1.3\times 10^{-15}$, which is negligible.
 Moreover, if $|I| = 500$, the upper bound is $1.2\times 10^{-93}$.
  If $|I|$ further increases, the upper bound of $\delta^{\prime}$ will keep decreasing. 
}

% we have
% \begin{equation}\label{eq:cdf_bio}
% \begin{aligned}
%     \delta^\prime <&\sum_{0}^{\ceil{2+\log\frac{4}{\delta}+\sqrt{4\log\frac{4}{\delta}+\left(\log\frac{4}{\delta}\right)^2}} }{|I|\choose i} \big(2(1-\pB)\big)^i \big(1-2(1-\pB)\big)^{|I|-i}  \\
%     &-\sum_{0}^{\floor{2\log\frac{4}{\delta}} }{|I|\choose i} \big(2(1-\pB)\big)^i \big(1-2(1-\pB)\big)^{|I|-i} 
% \end{aligned}
% \end{equation}

% as $|T_1|=s_1\sim \operatorname{Bin}(|I|,2(1-\pB))$.
% Let $\delta^\prime \triangleq  \Pr(0<|T_1|<\frac{1}{2}\left(\log\frac{4}{\delta}+\sqrt{\log\frac{4}{\delta}+4}\right)^2)$

(3) At last, when $\frac{|T_1|}{2}-\sqrt{\frac{|T_1|}{2}\log\frac{4}{\delta}}\geq 1$, which is further reduced to 
{\begin{equation*}
    |T_1|\geq 2+\log\frac{4}{\delta}+\sqrt{4\log\frac{4}{\delta}+\left(\log\frac{4}{\delta}\right)^2},
\end{equation*}}
then $\max\{1+\epsilon_{1,\mathsf{z}_1}, 1+\epsilon_{5,\mathsf{z}_1}, 1+\epsilon_{7,\mathsf{z}_1}, 1+\epsilon_{8,\mathsf{z}_1}\}  = 1+\epsilon_{7,\mathsf{z}_1}$.

Combining the results of (1)-(3), 
we have the DP guarantee, which is the worst case with
\begin{equation*}
    \begin{aligned}
     \epsilon_{\mathsf{z}_1} = \epsilon_{7,\mathsf{z}_1} =  & \frac{\frac{|T_1|}{2}+\sqrt{\frac{|T_1|}{2}\log\frac{4}{\delta}}}{\frac{|T_1|}{2}-\sqrt{\frac{|T_1|}{2}\log\frac{4}{\delta}}}, 
    \end{aligned}
\end{equation*}
%where $(*)$ holds as long as {\color{blue}$|T_1|\geq 18\log \frac{4}{\delta}$, which is guaranteed shown in Lemma 12}. 
which  completes the proof. % by setting $\epsilon = 1+ \sqrt{\frac{18\log\frac{4}{\delta}}{|T_1|}}.$
\end{IEEEproof}

\subsection{Proof of Lemma~\ref{claim:dp_amp_sub}}
\label{app:2nd-random}
For any neighboring data set $B\sim\widetilde B$, without loss of generality we denote the only different data pairs is   $b_j\neq \widetilde b_j$. 
% We aim to show that for all$\frac{\Pr(C_{s_1}(B)\in W)-\frac{\delta}{2}}{\Pr(C_{s_1}(\widetilde{B})\in W)}\leq 1+\left(1-2(1-\pB)\right) \sqrt{\frac{18\log\frac{4}{\delta}}{s_1}}$ holds.
Depending on whether $b_j$ and $\widetilde b_j$ belongs to the intersection or not, we need to consider the following 3 cases.
% \begin{equation*}
% \frac{\Pr(C_{s_1}(B)\in W)-\frac{\delta}{2}}{\Pr(C_{s_1}(\widetilde{B})\in W)}\leq 1+\left(1-2(1-\pB)\right) \sqrt{\frac{18\log\frac{4}{\delta}}{s_1}}.
% \end{equation*}

% First, we can have the derivation in (\ref{eq:part1}) (see Table \ref{table:claim_amp})

\noindent\textbf{Case a:} $b_j \notin I$ and $\widetilde b_j \notin \widetilde I$. In this case, $b_j\in B\setminus I$ and $\widetilde b_j\in \widetilde B\setminus \widetilde I$, which implies that $\Pr(C_{s_1}(B)\in W) = \Pr(C_{s_1}(\widetilde B)\in W)$ according to Algorithm~\ref{algo:decomp2}. Thus, $\frac{\Pr(C_{s_1}(B)\in W)-\frac{\delta}{2}}{\Pr(C_{s_1}(\widetilde{B})\in W)}\leq 1$  always holds. 

% Different with Algorithm~\ref{algo:decomp1} where $T_1$ is a constant subset of $I$, in Algorithm~\ref{algo:decomp2}, $\mathsf T_1$ is a random subset of $I$, Thus, we need to 

\noindent\textbf{Case b:} $b_j \in I$ and $\widetilde b_j \in \widetilde I$.
In this case, $b_j$ ($\widetilde b_j$) could  either belongs to $ T_1$ ($\widetilde T_1$) or $ I\setminus  T_1$ ($\widetilde I\setminus \widetilde T_1$ ).
Thus, we have Eq.~(\ref{eq:characteristic_part1}).
\begin{figure*}[!b]
% \centering
\normalsize
\setcounter{MYtempeqncnt}{\value{equation}}
\setcounter{equation}{35}
\hrulefill
\begin{equation}
\begin{small} 
\label{eq:characteristic_part1}
        \begin{aligned}
&\frac{\Pr[C_{s_1}(B)\in W|T_1]\Pr[T_1]-\frac{\delta}{2}}{\Pr[C_{s_1}(\widetilde{B})\in W|\widetilde T_1]\Pr[\widetilde T_1]}\\
= & \frac{  {n \choose s_1}^{-1}  \Big( \Pr\left[\left\{C_{s_1}(B)\in W\right\} \cap \left\{b_j\in T_1\right\} \right] 
    + \Pr\left[\left\{C_{s_1}(B)\in W\right\} \cap \left\{b_j\in I\setminus T_1\right\}    \right]  \Big)  -\frac{\delta}{2}}{  {n \choose s_1}^{-1}  \Big( \Pr\left[\left\{C_{s_1}(\widetilde{B})\in W\right\} \cap \left\{\widetilde b_j\in \widetilde{T}_1\right\} \right] + \Pr\left[\left\{C_{s_1}(\widetilde{B})\in W\right\} \cap \left\{\widetilde b_j\in \widetilde{I}\setminus \widetilde{T}_1\right\}\right] \Big) }\\
=& \frac{ {n \choose s_1}^{-1}  \Big(\Pr\left[\left\{C_{s_1}(B)\in W\right\} \Big| \left\{b_j\in T_1\right\} \right] 2(1-\pB) 
    + \Pr\left[\left\{C_{s_1}(B)\in W\right\} \Big| \left\{b_j\in I\setminus T_1\right\}    \right]  \left(1-2(1-\pB)\right)   -\frac{\delta}{2}     \Big)}{  {n \choose s_1}^{-1}  \Big(\Pr\left[\left\{C_{s_1}(\widetilde{B})\in W\right\} \Big| \left\{\widetilde b_j\in \widetilde{T}_1\right\} \right] 2(1-\pB) 
    + \Pr\left[\left\{C_{s_1}(\widetilde{B})\in W\right\} \Big| \left\{\widetilde b_j\in \widetilde{I}\setminus \widetilde{T}_1\right\}    \right]  \left(1-2(1-\pB)\right) \Big)  } <1,%\\
% =& \frac{ \gamma(B) 2(1-\pB) 
%     + \zeta(B)  \left(1-2(1-\pB)\right)  -\frac{\delta}{2}     }{  \gamma( B) 2(1-\pB) 
%     + \zeta(\widetilde B)  \left(1-2(1-\pB)\right)  }\\
    \end{aligned}
    \end{small}
\end{equation}

\end{figure*}
In Eq.~(\ref{eq:characteristic_part1}) the last inequality holds because $\Pr\left[\left\{C_{s_1}(B)\in W\right\} \Big| \left\{b_j\in T_1\right\} \right]= \Pr\left[\left\{C_{s_1}(\widetilde B)\in W\right\} \Big| \left\{\widetilde b_j\in \widetilde T_1\right\} \right]$ (c.f. \textbf{Case 9} in the proof of Lemma~\ref{claim:C_T_dp}) and $\Pr\left[\left\{C_{s_1}(B)\in W\right\} \Big| \left\{b_j\in I\setminus T_1\right\}    \right]  = \Pr\left[\left\{C_{s_1}(\widetilde B)\in W\right\} \Big| \left\{\widetilde b_j\in \widetilde I\setminus \widetilde T_1\right\}    \right]  $ (c.f. \textbf{Case 4} in the proof of Lemma~\ref{claim:C_T_dp}).

\noindent\textbf{Case c:} $b_j \in I$ and $\widetilde b_j \notin \widetilde I$ (i.e., $\widetilde b_j \in\widetilde B\setminus \widetilde I$).
Then we have Eq.~(\ref{eq:characteristic_part2}) as shown in the bottom of this page.
\begin{figure*}[!b]
\normalsize
\setcounter{MYtempeqncnt}{\value{equation}}
\setcounter{equation}{36}
\hrulefill
\begin{equation}
\begin{small} 
\label{eq:characteristic_part2}
   \begin{aligned}
   &     \frac{\Pr[C_{s_1}(B)\in W|T_1]\Pr[T_1]-\frac{\delta}{2}}{\Pr[C_{s_1}(\widetilde{B})\in W|\widetilde T_1]\Pr[\widetilde T_1]}\\
= & \frac{ {n \choose s_1}^{-1}\Big( \Pr\left[\left\{C_{s_1}(B)\in W\right\} \cap \left\{b_j\in T_1\right\} \right] 
    + \Pr\left[\left\{C_{s_1}(B)\in W\right\} \cap \left\{b_j\in I\setminus T_1\right\}    \right] \Big)   -\frac{\delta}{2}}{{n \choose s_1}^{-1} \Pr\left[\left\{C_{s_1}(\widetilde{B})\in W\Big| \widetilde b_j\notin \widetilde I\right\}   \right] }\\
    =& \frac{ {n \choose s_1}^{-1}\Big(\Pr\left[\left\{C_{s_1}(B)\in W\right\} \Big| \left\{b_j\in T_1\right\} \right] 2(1-\pB) 
    + \Pr\left[\left\{C_{s_1}(B)\in W\right\} \Big| \left\{b_j\in I\setminus T_1\right\}    \right]  \left(1-2(1-\pB)\right) \Big)  -\frac{\delta}{2}      }{ {n \choose s_1}^{-1}  \Pr\left[\left\{C_{s_1}(\widetilde{B})\in W\Big| \widetilde b_j\in \widetilde B\setminus \widetilde I\right\}   \right] } \\
\stackrel{(a)}= & \frac{ {n \choose s_1}^{-1}\Big(\gamma(B) 2(1-\pB) 
    + \zeta(B) \left(1-2(1-\pB)\right) \Big)  -\frac{\delta}{2}   }{ {n \choose s_1}^{-1} \kappa (\widetilde B)}\\
    \stackrel{(b)} < &\frac{ {n \choose s_1}^{-1} \Big(\left((1+\epsilon_{8,\mathsf{z_1}})  \kappa (\widetilde B)+ \frac{\delta}{2} \right)2(1-\pB) 
    + \left((1+\epsilon_{5,\mathsf{z}_1})\kappa (\widetilde B) +\frac{\delta}{2} \right)\left(1-2(1-\pB)\right) \Big)  -\frac{\delta}{2}      }{ {n \choose s_1}^{-1}\kappa (\widetilde B)} \\
 < &\frac{{n \choose s_1}^{-1} \Big( (1+\epsilon_{8,\mathsf{z}_1})  \kappa (\widetilde B) 2(1-\pB) 
    +  (1+\epsilon_5)\kappa (\widetilde B)  \left(1-2(1-\pB)\right)  \Big)   }{{n \choose s_1}^{-1}\kappa (\widetilde B)}  = 1+2(1-\pB) \epsilon_{8,\mathsf{z}_1} + \left(1-2(1-\pB)\right)\epsilon_{5,\mathsf{z}_1}, % \epsilon_5 \leq 1+\max\{\epsilon_5,\epsilon_8\},
    \end{aligned}
    \end{small}
\end{equation}
\end{figure*}
In Eq~(\ref{eq:characteristic_part2}), (a) can be achieved by defining
\begin{equation}
% \begin{aligned}
\gamma(B) \triangleq \Pr\left[\left\{C_{s_1}(B)\in W\right\} \Big| \left\{b_j\in T_1\right\} \right]
\end{equation} 
\begin{equation}
   \zeta(B) 
   \triangleq \Pr\left[\left\{C_{s_1}(B)\in W\right\} \Big| \left\{b_j\in I\setminus T_1\right\} \right],  \end{equation}
   \begin{equation} \kappa(\widetilde B) \triangleq  \Pr\left[\left\{C_{s_1}(\widetilde B)\in W\right\} \Big| \left\{\widetilde b_j\in \widetilde B\setminus \widetilde I\right\} \right] 
\end{equation}
and $(b)$ is because
$\gamma(B)\leq (1+\epsilon_{8,\mathsf{z}_1})\kappa(\widetilde B)+\frac{\delta}{2}$ (c.f. \textbf{Case 8} in the proof of Lemma~\ref{claim:C_T_dp}) and $\zeta(B)\leq (1+\epsilon_{5,\mathsf{z}_1})\kappa(\widetilde B)+\frac{\delta}{2}$ (c.f. \textbf{Case 5} in the proof of Lemma~\ref{claim:C_T_dp}). 
{Furthermore, one can verify that $\epsilon_{5,\mathsf{z}_1}>\epsilon_{8,\mathsf{z}_1}$ for \textbf{Scenario iii} in the proof of Lemma \ref{claim:C_T_dp}, 
% if $|T_1|>2\log\frac{4}{\delta}$, which naturally holds when $s_1>\max\left\{2+\log\frac{4}{\delta}+\sqrt{4\log\frac{4}{\delta}+\left(\log\frac{4}{\delta}\right)^2},18\log\frac{4}{\delta}\right\}$, 
thus \textbf{Case c} is upper bounded by $1+\epsilon_{5,\mathsf{z}_1}$ with $s_1\leftarrow \mathsf |T_1|$, and 
$$\epsilon_{\mathsf z_{2}} = \frac{\frac{s_1}{2}+\sqrt{\frac{s_1}{2}\log\frac{4}{\delta}}+1}{\frac{s_1}{2}-\sqrt{\frac{s_1}{2}\log\frac{4}{\delta}}} - 1.$$}

\subsection{Proof of Theorem~\ref{claim:dp_I_int_AB}}
\label{app:A-DP}

\begin{IEEEproof}
According to Line~1 in Algorithm~\ref{algo:observe_card3},
$s_1=|T_1|$  is the summation of $|I|$ Bernoulli trials with parameter $2(1-\pB)$. Applying the Hoeffding's inequality to the summation of these r.v., we have
\begin{equation*}
    \Pr\left[\Big|\mathsf s_1-2(1-\pB)|I|\Big|<\sqrt{\frac{|I|}{2}\log\frac{2}{\dB}}\right]>1-\delta_b,
\end{equation*}
which suggests that {$ \Pr\left[\mathsf s_1\in \mathcal S^c\right]\leq\frac{\dB}{2}$}
with  $\mathcal S^c$ the complementary set of $\mathcal S$ defined by
\begin{equation}
\label{eq:T1-range}
\mathcal S\triangleq
\left\{\mathsf s_1: \mathsf s_1\geq 2(1-\pB)|I|-\sqrt{\frac{|I|}{2}\log\frac{2}{\dB}}\right\}.
\end{equation}
Then, for any pair of neighboring data sets $B\sim\widetilde{B}$ and any subset $W\subseteq [n]$, the PMF of   $ \mathsf z_1$ is 
\begin{equation}\label{eq:random_s}
    \begin{aligned}
    &\Pr[\mathsf z_1\in W]\\
    = &\Pr\left[\mathsf z_1\in W\cap \mathsf s_1\in \mathcal S\right]  +\Pr\left[\mathsf z_1\in W\cap \mathsf s_1\in \mathcal S^c\right]  \\
\leq & \Pr\left[\mathsf z_1\in W\cap \mathsf s_1\in \mathcal S\right] +\frac{\dB}{2}\\
= & \sum_{s_1\in \mathcal S}\Pr\left[C_{s_1}\in W | \mathsf s_1\right] \Pr[\mathsf{s_1}=s_1]+\frac{\dB}{2}.
    \end{aligned}
\end{equation}
%Denote $\epsilon_{\#} = \max\{\epsilon_5,\epsilon_8\}$, and
% According to Lemma \ref{claim:d\pAmp_sub}, when  {\color{blue}$s_1\geq 2(1-\pB)|I|-\sqrt{\frac{|I|}{2}\log\frac{2}{\delta}}>\max\left\{2+\log\frac{4}{\delta}+\sqrt{4\log\frac{4}{\delta}+\left(\log\frac{4}{\delta}\right)^2},18\log\frac{4}{\delta}\right\}$}, i.e.,
% \begin{equation}\label{eq:p_epsilon}
% \begin{aligned}
%     p<\min\Bigg\{&1-\frac{\sqrt{\frac{|I|}{2}\log\frac{4}{\delta}}+18\log\frac{4}{\delta}}{2|I|},\\
%     &1-\frac{\sqrt{\frac{|I|}{2}\log\frac{4}{\delta}}}{2|I|}-\frac{2+\log\frac{4}{\delta}+\sqrt{4\log\frac{4}{\delta}+\left(\log\frac{4}{\delta}\right)^2}}{2|I|}\Bigg\},
%     \end{aligned}
% \end{equation}
%$p<1-\frac{\sqrt{\frac{|I|}{2}\log\frac{4}{\delta}}+18\log\frac{4}{\delta}}{2|I|}$, 
By substituting the result of Lemma~\ref{claim:dp_amp_sub} into the above inequality, we obtain
% \begin{equation*}
%      \Pr\left[C_{s_1}\in W  \right]\leq e^{\epsilon_{5}} \Pr\left[C_{s_1}(\widetilde{B})\in W \right]+\frac{\delta}{2}.
% \end{equation*}
% Thus, we have
\begin{equation*}
\begin{aligned}
&\Pr[\mathsf z_1\in W]\\ \leq&\left(\sum_{s_1\in \mathcal S_b}\left(e^{\epsilon_{5,\mathsf{z}_1}}\Pr\left[C_{s_1}(\widetilde{B})\in W |s_1 \right]+\frac{\delta}{2}\right) \Pr[\mathsf{s_1}=s_1]\right)+\frac{\delta}{2}\\
\leq& \left(\sum_{s_1\in \mathcal S_b}e^{\epsilon_{5,\mathsf{z}_1}}\Pr\left[C_{s_1}(\widetilde{B})\in W |s_1\right] \Pr[\mathsf{s_1}=s_1]\right)+ \delta\\
% \leq& \left(\sum_{s_1\geq2(1-\pB)|I|-\sqrt{\frac{|I|}{2}\log\frac{2}{\delta}}}e^{\epsilon_{\#}}\Pr\left[C_{s_1}(\widetilde{B})\in W \right] \Pr[\mathsf{s_1}=s_1]\right)+ \delta\\
\leq& \max_{s_1\in \mathcal S_b}e^{\epsilon_{5,\mathsf{z}_1}}\Pr\left[C_{s_1}(\widetilde{B})\in W|s_1 \right]   + \delta
\end{aligned}
\end{equation*}
% Since $|T_1|$ increases with $s$; the larger the size $T$ (sampled data from Bob) the larger the overlapping between $T$ and $I$, thus, $\epsilon_{\#}$ decreases with $s$. As a result, the above achieves the maximum at the smallest cardinality of $T$ (denoted as $T_{\min}$, $|T_{\min}| = \lambda-\sqrt{\frac{n}{2}\log\frac{4}{\delta}}$). Hence,  
To guarantee that $\epsilon_{5,\mathsf{z}_1}$ in  (\ref{eq:eps5-2}), which we repeat here
% \begin{equation}
$
\epsilon_{5,\mathsf{z}_1} = \frac{2\sqrt{\frac{|T_1|}{2}\log\frac{4}{\delta_B}}+1}{\frac{|T_1|}{2}-\sqrt{\frac{|T_1|}{2}\log\frac{4}{\delta_B}}}
$,
% \end{equation}
be positive, it is required to ensure the denominator be positive.  { Solving this positive denominator inequality, we obtain the constraint for $T_1$ as below}
\begin{equation}
\label{eq:e5-denomitor}
   |T_1|>2\log\frac{4}{\dB}. 
\end{equation}
Moreover,  when $|T_1|>2\log\frac{4}{\delta}$, $\epsilon_{5,\mathsf{z}_1}$ is a monotonic decreasing function w.r.t. $|T_1|$ (The proof please refer to that in to proof the monotonic decreasing property of $\eB$ w.r.t. $T$ by simply replacing $T/2$ with $t$). Thus $\epsilon_{5,\mathsf{z}_1}$ achieves the maximum at the smallest $|T_1|$.
By combining (\ref{eq:T1-range}) and (\ref{eq:e5-denomitor}), {one sufficient condition for }
\begin{equation}
\label{eq:eB-1}
\eB=\epsilon_{5,\mathsf{z}_1} = \frac{2\sqrt{\frac{|T_1|}{2}\log\frac{4}{\dB}}+1}{\frac{|T_1|}{2}-\sqrt{\frac{|T_1|}{2}\log\frac{4}{\dB}}}.
\end{equation}
is given by 
\begin{equation}
\label{eq:region}
 |T_1|\geq  2(1-\pB)|I|-\sqrt{\frac{|I|}{2}\log\frac{2}{\dB}}> 2\log\frac{4}{\dB}.
\end{equation}
Solving the  inequalities in (\ref{eq:region}) w.r.t. $|I|$, the  sufficient condition for $\eB$ in (\ref{eq:eB-1}) is further 
reduced  to 
\begin{equation}
\begin{cases}
 |T_1|\geq  2(1-\pB)|I|-\sqrt{\frac{|I|}{2}\log\frac{2}{\dB}},\\
|I|>
\frac{\left(\sqrt{\frac{1}{2}\log\frac{2}{\dB}} + 
\sqrt{\frac{1}{2}\log\frac{2}{\dB}+16(1-\pB)\log\frac{4}{\dB}}\right)^2 }
{16(1-\pB)^2}
\triangleq L.
\end{cases}
\end{equation}
Since $\eB$ is a monotonic decreasing function w.r.t. $T_1$ (as shown in Proposition ??), the largest $\eB$ is obtained {when $|T_1|=  2(1-\pB)|I|-\sqrt{\frac{|I|}{2}\log\frac{2}{\dB}}$}, which leads to
% \begin{equation*}
% \begin{aligned}
%     \Pr[\mathsf z_1 (B)\in W]\leq &e^{\epsilon_B}\Pr[\mathsf z_1 (\widetilde{B})\in W] +  \dB.
% \end{aligned}
% \end{equation*}
% where
\begin{equation}
\label{eq:epsilonB}
\epsilon_B = \frac{2\sqrt{t\log\frac{4}{\dB}}+1}{t-\sqrt{t\log\frac{4}{\dB}}}
\end{equation}
with $t \triangleq \frac{|T_1|}{2}= (1-\pB)|I|-\sqrt{\frac{|I|}{8}\log\frac{2}{\dB}}
> 2\log\frac{4}{\delta_B}$ and 
$|I|>
\frac{\left(\sqrt{\frac{1}{2}\log\frac{2}{\dB}} + 
\sqrt{\frac{1}{2}\log\frac{2}{\dB}+16(1-\pB)\log\frac{4}{\dB}}\right)^2 }
{16(1-\pB)^2}.$

\end{IEEEproof}

\subsection{Proof of Theorem~\ref{lemma:region}}
\label{app:rr}
\begin{proof}
Bob's view is $\eA$-DP for $A$   if for neighboring data sets $A\sim A^\prime$ and all $k, j,j^{\prime}\in\bin$, we have

\begin{equation}
\begin{aligned}
&\frac{\Pr(\mathcal M_A(A, B) \in F)}{\Pr(\mathcal M_A(A^{\prime}, B) \in F)}\\
=&
\frac{\prod_{a_{\ell}\in A}
\Pr\left(X_{i}=k \mid {\mathbbm 1}_{A\cap  B}(a_\ell)=j\right)
% \Pr\left( {\mathbbm 1}_{A\cap  B}(a_\ell)=j\right)
}
{\prod_{ a_{\ell}\in A^{\prime}} 
\Pr\left(X_{i}=k \mid {\mathbbm 1}_{A^\prime \cap  B}(a_{\ell})=j^{\prime}\right)}
\\
\leq & e^{\eA},
\end{aligned}
\end{equation}
for arbitrary $F\subseteq \{0,1\}^{\ast\times m}$.
Fix neighboring data sets $A\sim A^{\prime}$ with the only different elements with index $\ell$, where $a_{\ell}\neq a^{\prime}_{\ell}$ 
 Then the above $\epsilon$-DP definition can be further simplified as 
\begin{equation}
\label{dp-rr}
\frac{\Pr(\mathcal M(A, B) \in F)}{\Pr(\mathcal M(A^{\prime}, B) \in F)}
=
\frac{ 
\Pr\left(X_{\ell}=i \mid {\mathbbm 1}_{A\cap  B}(a_{\ell})=j\right)}
{
\Pr\left(X_{\ell}=i \mid {\mathbbm 1}_{A^{\prime}\cap B}(a^{\prime}_{\ell})=t\right)} \leq e^{\eA}.
\end{equation}
By substituting (\ref{eq:TP})-(\ref{eq:FN})  into~(\ref{dp-rr}), we have the following valid region
to achieve $\eA$-DP:
\begin{equation}
\left\{
             \begin{array}{lr}
\pA\leq e^{\eA}q & \\ 
1-q  \leq e^{\eA}\left(1-\pB\right) &\\ 
q\leq e^{\eA} \pA &\\
1-\pA\leq e^{\eA}(1-q) &\\
0\leq \pA, q \leq 1 &
             \end{array}.
\right .
\end{equation}
In our DP$^2$SI, under the same privacy budget $\eA$, we prefer a larger $p$ and smaller $q$ corresponding the TN and FN case. With the assumption that $p\geq q$, we simplify the above valid region as $\mathcal R$ shown in (\ref{eq:epsilon-region}), which completes the proof. 
\end{proof}

\subsection{Proof of Lemma~\ref{optimal_p_q}}
\label{app:precision-recall}
\begin{proof}
We first maximize the expected precision under the DP  constraint, i.e.,
\begin{equation}\label{precision}
\begin{array}{rrclcl}
\displaystyle \max_{\pA, q} & \frac{\pA|A\cap B| }
{
q|B_{\mathsf{sub}}\setminus A| + \pA| A\cap B|}\\
\textrm{s.t.} & \pA, q\in \mathcal{R}\quad \text{in} \quad  (\ref{eq:epsilon-region}).
\end{array}
\end{equation}
Note that (\ref{precision}) is  monotonic increasing w.r.t. $\pA/q$.
With the valid $\pA$ and $q$ defined by $\mathcal R$ in (\ref{eq:epsilon-region}), we have the optimal $\pA/q$ is $e^{\eA}$ with the optimal precision 
$\frac{|A\cap B| }
{
e^{-\eA}|B_{\mathsf{sub}}\setminus A| + | A\cap B|}$.
Similarly, we obtain the maximum (\ref{eq:recall}) is obtained with the same $\pA/q = e^{\eA}$.
We therefore obtain $\pA^{\ast}= \frac{e^{\epsilon}}{1+e^{\epsilon}}$  and $q^{\ast} = \frac{1}{1+e^{\epsilon}}$.
\end{proof}

\subsection{Proof of Proposition~\ref{prop:increasing}}
\label{app:I-condition}

\begin{proof}
We first compute the monotonically increasing region for function $t$. Setting $\frac{\partial t}{\partial |I|}>0$, we obtain
$$\frac{\partial t}{\partial |I|} =  1-\pB - \frac{1}{4}\sqrt{\frac{1}{2}\log\frac{2}{\dB}}\frac{1}{\sqrt{|I|}}>0,$$ which is equivalent to
\begin{equation}
    |I|>\frac{1}{32(1-\pB)^2}\log\frac{2}{\dB}.
\end{equation}
Moreover, setting $t = (1-\pB)|I|-\sqrt{\frac{|I|}{8}\log\frac{2}{\dB}}>0$
we obtain
$|I|>\frac{1}{8(1-\pB)^2}\log\frac{2}{\dB}$.
Thus, $t$ is  monotonically increasing for 
$|I|>\frac{1}{8(1-\pB)^2}\log\frac{2}{\dB}$.
\end{proof}

\subsection{Proof of Proposition~\ref{prop:decreasing}}
\begin{proof}
We first reformulate $\eB$ by $\eB = \frac{2\sqrt{t\log\frac{4}{\dB}}+1}{t-\sqrt{t\log\frac{4}{\dB}}}  = \frac{2\sqrt{\log\frac{4}{\dB}}+\frac{1}{\sqrt{t}}}{\sqrt{t}-\sqrt{\log\frac{4}{\dB}}}$
and define $k \triangleq \sqrt{\log\frac{4}{\dB}}$.
Hence $\eB = \frac{2k+\frac{1}{\sqrt{t}}}{\sqrt{t}-k}$.
The first order derivative of $\eB$, therefore, is given by
$$\frac{\partial g(t)}{\partial t} = \frac{\frac{k}{2}t^{-3/2}-kt^{-1/2}-t^{-1}}{(\sqrt{t}-k)^2} = \frac{\frac{k}{2}t^{-1}-k-t^{-1/2}}{\sqrt{t}(\sqrt{t}-k)^2}.$$
We then obtain that
$\frac{\partial g(t)}{\partial t}<0$ when $\frac{1}{\sqrt{t}}\in (0,\frac{1+\sqrt{1+2k^2}}{k})$. Thus, if $\frac{1}{\sqrt{t}}\in (0,\frac{1}{k})$, i.e., $t>k^2 = \log \frac{4}{\dB}$, $\frac{\partial g(t)}{\partial t}<0$ always holds, which completes the proof. 
\end{proof}

%%%%%%%%%%%%%%%%%%%%%%%%%%%%%%%%%%%%%%%%%%%%%%%%%%%%%%%%%%%

\bibliographystyle{plain}
\bibliography{Jian}

% \begin{figure*}[tp]
%     \centering
%     \label{Alg:DPSetIntersection-temp}
%     \input{protocols/dp_psi_stat_protocol}
%     \caption{DP Budget for PSI-Statistics}
% \end{figure*}
% \section{Appendix A}
% \label{FirstAppendix}

% \section{Appendix B}
% \label{FirstSubsectionAppendix}
\end{document}